\documentclass[twocolumn,secnumarabic, amsmath,amssymb, nobibnotes, aps, prd,floatfix,showkeys, showpacs]{revtex4-1}
\usepackage{siunitx}
\usepackage{graphicx}
\usepackage{lipsum}
\usepackage{pgf, tikz}
\usepackage[pdfusetitle]{hyperref}
\usepackage{color}
\usepackage[percent]{overpic}
\usepackage[normalem]{ulem} 

\hypersetup{pdfinfo={
Title={Boundary-induced nucleation control: a theoretical perspective},
Author={O. Buller, H. Wang, W.C. Wang, L. F. Chi, A. Heuer},
Subject={Structure formation},
Keywords={organic molecules, surface diffusion, aggregate, nucleation},
}}
\newcommand{\kbT}[0]{ k_\mathrm{B}T}
\newcommand{\avN}[0]{\langle N \rangle}

\newcommand{\lowt}[1]{ _{\text{#1}}}
\newcommand{\avr}[1]{\left \langle #1 \right \rangle}
\newcommand{\figref}[1]{Fig.~#1}
\newcommand{\eqaref}[1]{Eq.~#1}
\newcommand{\secref}[1]{Sect.~#1}
\newcommand{\tabref}[1]{Table~#1}

\newcommand{\subfigimg}[3][,]{%
  \setbox1=\hbox{\includegraphics[#1]{#3}}
  \leavevmode\rlap{\usebox1}
  \rlap{\hspace*{-10pt}\raisebox{\dimexpr\ht1-1\baselineskip}{#2}}
  \phantom{\usebox1}
}

\begin{document}

\title{Boundary-induced nucleation control: A theoretical perspective}%

\author{Oleg Buller$^1$}
\email[ To whom correspondence should be addressed: ]{o.buller@wwu.de}
\author{Hong Wang$^2$}
\author{Wenchong Wang$^{2}$}
\author{Lifeng Chi$^{2,3}$}%
\author{Andreas Heuer$^1$}
\affiliation{1 Institute for Physical Chemistry, University of M\"unster, Correnstr. 28/30, 48149 M\"unster, Germany}
\affiliation{2 Physikalisches Institut and Center for Nanotechnology (CeNTech), University of M\"unster, 48149 M{\"u}nster, Germany}
\affiliation{3 Jiangsu Key Laboratory for Carbon-based Functional Materials \& Devices Collaborative,
 Institute of Functional Nano \& Soft Materials (FUNSOM) and Collaborative Innovation Center of
Suzhou Nano Science and Technology, Soochow University, Suzhou 215123, P. R. China}

\begin{abstract}
The pre-patterning of a substrate to create energetically more attractive or repulsive regions allows one to generate a variety
of structures in physical vapor deposition experiments.
A particular interesting structure is generated if the energetically
attractive region is forming a rectangular grid. For specific combinations of the particle
flux, the substrate temperature and the lattice size it is possible to generate exactly one
cluster per cell, giving rise to nucleation control. Here, we show that the experimental
observations of nucleation control can be very well understood from a theoretical perspective.
For this purpose we perform, on the one hand, kinetic Monte Carlo simulations and, on the other
hand, use analytical scaling arguments to rationalize the observed behavior.
For several observables, characterizing
nucleation control, a very good agreement is found between experiment and theory. This underlines
the generality of the presented mechanism to control the deposition of material by manipulation
of the direct environment.
\end{abstract}


\maketitle

\section{Introduction}\label{sec:intro}

With physical vapor deposition (PVD) material can be condensed extensively onto a substrate, yielding
films with desired thickness on the desired length scales \cite{Mahan2000physical, Mattox2010handbook}.
Atoms and molecules, also as multiple component systems, can be deposited in this way.

During the deposition the nucleation process is of major impotence, therefore
a critical number of particles, i.e. atoms or molecules, has to come
together to form stable clusters \cite{islands_mounds_atoms,Evans2006morphological}. Before nucleation,
on a homogeneous substrate the ad-atom density is uniform and increases linearly, in the scope of a mean-field
description. The location of the first nucleus is totally random and caused by local ad-atom density fluctuations,
afterwards the nucleation site becomes an absorber for the ad-atoms on the surface.

In order to place material heterogeneously on a desired position,  the method of template-directed nucleation was introduced
\cite{Niu2006level, Nurminen200063, Choongseop199873, Wang200798, Kalischewski200878, Hu2008101}.
Here, the substrate is first chemically pre-patterned with a material, which for the deposited particles
is energetically more attractive than the substrate. This procedure is typically performed by lithographic
methods and applied on the (high) nano and micrometer scale \cite{Wang200798}.
Within a specific range of $D/F$, where $D$ is the surface diffusion coefficient and $F$ the
particle flux of the depositing material,
nucleation takes place only on the pre-defined positions, i.e. on the pre-pattern. This configuration corresponds
to the equilibrium state of the system and is realized in the limit of small flux.

In a different way the ''pre-patterning''  can be created by Moire structures, for instance by having
a monolayer of graphene on a transition metal substrate \cite{Engstfeld2012directed, Diaye2006twidimensional}. Through the
mismatch of lattices a periodically modulated structure is created, with a periodic variation of the binding-
and transition energies on the graphene layer. A regular formation of (nano-) clusters in the sub-monolayer regime
can be obtained by thermodynamic and kinetic effects \cite{Niu2006level}, due to the heterogeneous substrate field structure.
In this context this process is often referred by the term directed self-assembly.

The detailed mechanisms on the surface during PVD are rather complex \cite{Gadzuk2008fundamental,Kleyn2008basic} and
cover different time and space scales, from typical atomic vibrations (\SI{e-13}{\s}) to growth processes on the
mesoscopic scale with a duration
of minutes or hours. In order to overcome this complexity and focus on key relations and mechanisms of the growth process,
mean-field and coarse-grained approaches were introduced, which are formulated on the time scale of the transition processes.
With the mean-field rate equations, describing the ad-atom
and island densities with coupled ordinary differential equations, basic scaling relations of deposition growth
can be elucidated \cite{Venables1973rate,Hanbuecken1984399,islands_mounds_atoms,Evans2006morphological}. In order to capture the spatial
resolution, stochastic/kinetic Monte Carlo methods are very powerful \cite{Kotrla1996numerical,levi97:_theor,Jonsson2000theoretical, Voter2007}.
In this context there also exist hybrid models \cite{Petersen2001level,Ratsch2002level} that combine the mean-field with
the atomistic perspective.
Other related approaches are phase-field models \cite{Yu2004phase},
geometry based simulations \cite{Li2003geometry} or the Quasicontinuum Monte Carlo \cite{Russo2004quasicontinuum}.
The success of these methods lies in describing accurately  the main aspects of the structural
surface growth during PVD, namely the diffusion and nucleation processes.

In this paper we deal with a nucleation control method where the nucleation site is spatially
well-defined but which does not rely on
the presence of a pre-pattern at this location. Rather the spatial position is controlled through the boundaries,
given by an ad-atom adsorbing grid .  This method exploits the maximum of the ad-atom distribution between
sinks and the existence of a critical density where nucleation commences \cite{paperZero, Ranguelov200775}.
In this way, very regular structures can be generated , cf. \figref{\ref{fig:exp_fig}} left.  Note, that the mechanism
is kinetically driven and the central nucleus is placed directly on the substrate. In contrast to template-directed
nucleation this mechanism to generated ordered structures is a non-equilibrium process, which
sensitively depends on an appropriate choice of the ratio $D/F$, for a fixed grid size $L$.

The mechanism of that approach is explored based on the previous results and
the model system, put forward in \cite{paperZero}. The quality of the used lattice
kinetic Monte Carlo (KMC) model is  verified through direct comparison with
the experimental results for different observables which characterize properties of the
nuclei and in particular the nucleation control, i.e. the presence of one nucleus \textit{per} grid cell.
Furthermore, the dependence on the external flux of this particular structure is analyzed analytically based on the mean-field
rate equations; a scaling relation is derived and verified with KMC simulations.
Due to the agreement of the experimental and simulation outcome
as well as the analytical scaling results, we argue that the mechanism
of indirect nucleation control is of general nature and can be extended to a wide variety of scenarios.

The structure of this paper is as follows. In \secref{\ref{sec:exper-backgr}} we recap
experimental results of cluster formation followed by the simulation setup in \secref{\ref{sec:sim_setput}}.
Afterwards the scaling relation from mean-field rate equation approach is derived in
\secref{\ref{sec:theo_sing_cluster}} and
the qualitative behavior of cluster growth presented in \secref{\ref{sec:quanlit_behav}}. Nucleation- and position control is
discussed in \secref{\ref{sec:nuc_and_pos_control}}. The flux dependence of the nucleation control is
presented in  \secref{\ref{sec:flux_dep}}, the influence of the diffusion properties and scaling of
multiple islands in \secref{\ref{sec:nuc_to_std_nuc}}. Finally, we conclude with a discussion of our results
in \secref{\ref{sec:discussion}}.

\section{Experimental background}\label{sec:exper-backgr}

PVD methods are suited very well to deposit complex organic molecules, which are increasingly
used in microelectronics, due to the functionality and efficiency they provide.
Therefore, this type of molecules is of great interest from the experimental point of view
\cite{Forrest199797,Briseno2005patterned,Sundar2004elastomeric,Lucas201261}.
Here we briefly refer to the experimental set up. In general the experiments are conducted
in the same way as in \cite{paperZero}.

For deposition the functional molecule N,N'-bis(1-naphthyl)-N,N'-diphenyl-1,1'-biphenyl'-4,4'-diamine
(NPB, a molecule widely used for organic light emitting diodes) has been used  \cite{Forsythe199873}.
This molecule was deposited on a silica surface. To create an attractive region for the molecules,
a gold grid was put on the surface by standard beam lithography.
The gold grid is acting as an adsorbing environment for the central area of the squares.

\begin{figure}[t]
  \centering
  \subfigimg[width=0.45\textwidth]{a)}{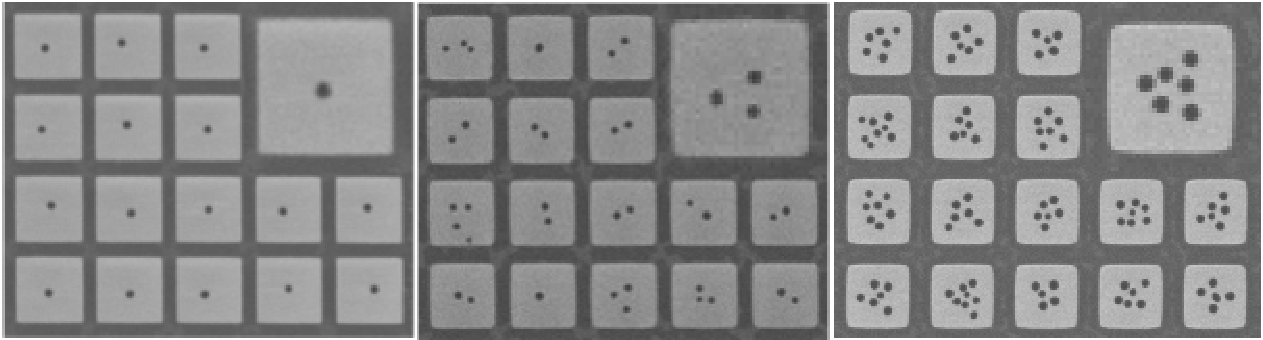}\\
  \subfigimg[width=0.45\textwidth]{b)}{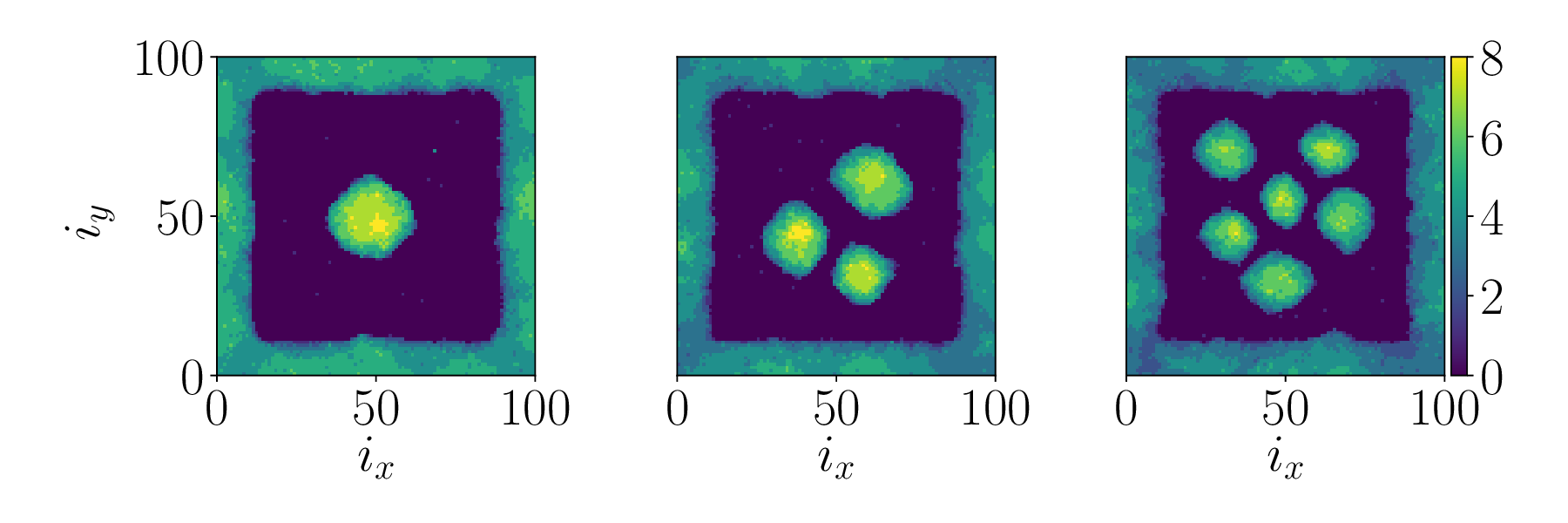}
  \caption{\label{fig:exp_fig} a) SEM pictures of formations on a grid pre-patternd surface with
    $L = \SI{4.0}{\um}$ and flux variation \SI{0.042}{\nm\per\minute},
    \SI{0.066}{\nm\per\minute}, \SI{0.085}{\nm\per\minute} from left to right. The right top
    corner shows the magnified view of one cell, respectively.
    b) The height profile of a similar result from the KMC model for the grid size of $80a$. The flux
    is varied with $5.85 \times 10^{-6}/(a^2 \Delta t)$, $1.40 \times 10^{-5}/(a^2 \Delta t)$
    and $2.60 \times 10^{-5}/(a^2 \Delta t)$ on going from left to right. The color-coding for the
    height is given on the right.
   }
\end{figure}

In \figref{\ref{fig:exp_fig}} a) the scanning electron microscope (SEM) images are shown of the resulting
structure for a gold array distance of $L=\SI{4.0}{\um}$. The flux was increased on going from left to right.
In case of the lowest flux $F= \SI{0.042}{\nm\per\minute}$, shown in \figref{\ref{fig:exp_fig}} a) left,
one cluster is created with a well-defined
position in the center of the cell. Due to the regularity of the grid the clusters also form a regular structure, i.e. show
position control. With increasing the flux, the number of clusters increases, i.e. nucleation control is no longer present.
In the same way it works for other molecule types \cite{paperZero}. To prove the generality of the mechanism,
a \emph{solid-on-solid} model \cite{Biehl2005lattice} is used,
which only provides the basic mechanism of surface growth, i.e. a particle flux,
surface diffusion and nucleation through parametrized interaction strength.
Note, that within this model the deposit molecule is reduced to a point
on a lattice with isotropic interaction and diffusion properties. An impression
of the result from the simulations is shown in \figref{\ref{fig:exp_fig}} b) with a
similar scaling for the island number by varying the flux.
The KMC model described in the following section.

\section{Simulation Setup}
\label{sec:sim_setput}
For the simulations a lattice gas model is used on a three dimensional cubic lattice with
a node distance of $a$ as described in \cite{Biehl2005lattice,paperZero}.
Every lattice site ${\bf i}=\{i_x, i_y, i_z\}$ can be occupied
by one of the three different particle types: deposited particles $p_{\bf i}$, substrate sites
$s_{\bf i}$ and the more attractive pre-pattern sites $g_{\bf i}$ (in the experiment represented by gold). The lattice site
${\bf i}$ is either filled or empty. As we are only interested in the dynamics of the deposited particles the
Hamiltonian can be written as
\begin{equation}
\label{hamiltonian}
\begin{split}
  H = -\varepsilon_{pp} \; \frac 1 2 \sum_{\bf i,j} f(r_{\bf ij}) p_{\bf i} p_{\bf j} \qquad \\
  - \varepsilon_{pg} \; \frac 1 2 \sum_{\bf i,j} f(r_{\bf ij}) p_{\bf i} g_{\bf j} -
    \varepsilon_{ps} \; \frac 1 2 \sum_{\bf i,j} f(r_{\bf ij}) p_{\bf i} s_{\bf j},
\end{split}
\end{equation}
where the $\varepsilon_{xy} \, (x,y \in \{ p,g,s \})$ are the interaction
parameters. The distance scaling function $f(r_{\bf ij})$ depends on
the distance $r_{\bf ij}$ between the lattice sites on ${\bf i}$ and ${\bf j}$ and is
defined as follows:
  \begin{center}
    \begin{tabular}{l|c|c|c|c|c}
      \toprule
      $r_{\bf ij}$& 0.0 & 1.0 & $\sqrt 2$ & $\sqrt 3$ & $ > \sqrt 3$ \\ \hline
      $f(r_{\bf ij})$       & 0.0 & 1.0 & 1.0       & 0.5 & 0.0  \\ \hline
    \end{tabular}
  \end{center}
The interaction range up to the third nearest neighbors (common corner) is taken into account.
The interaction parameters are the same as in \cite{paperZero}, $\epsilon_{pg}/\kbT = 1.3$, $\epsilon_{pp}/\kbT = 1.0$  and
$\epsilon_{ps}/\kbT = 0.3$, where $k_{\mathrm{B}}$ is the Boltzmann constant and $T$ the temperature.
The necessary conditions for the boundary-induced nucleation are an ad-atom absorbing
environment and the island growth regime on the substrate. The absorbing environment
is here realized by a energetically more attractive grid pre-pattern, i.e. $\epsilon_{pg} > \epsilon_{ps}$.
In this way, the boundary of the pre-pattern represents a sink for the central
region. As in this case attaching to the pre-pattern is associated with an energy gain and the detachment with an energy
barrier, therefore the situation is similar to the process of aggregation to a nucleus.

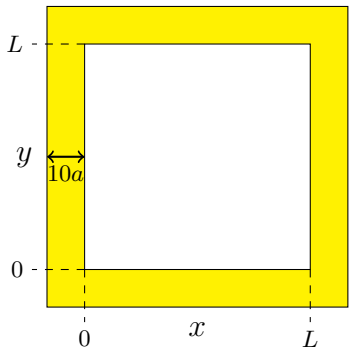
\begin{figure}[t]
  \centering
  \begin{tikzpicture}
    \coordinate (A) at (0.5,2.0);
    \coordinate (B) at (3.5,2.0);
    \draw [fill=yellow] (0.0,0.0) rectangle (4.0,4.0);
    \node[thick] at (2.0,-0.3) {\large{$x$}};
    \node[thick] at (-0.3,2.0) {\large{$y$}};
    \draw [fill=white] (0.5,0.5) rectangle (3.5,3.5);
    \draw[dashed] (0.5,0.5) -- (0.5,-0.2) node[below] {$0$};
    \draw[dashed] (3.5,0.5) --  (3.5,-0.2) node[below] {$L$};
    \draw[dashed] (0.5,0.5) -- (-0.2, 0.5) node[left] {$0$};
    \draw[dashed] (0.5,3.5) --  (-0.2, 3.5) node[left] {$L$};
    \draw[<->, thick] (0.0,2.0)  -- node[below] {$10a$} (A);
  \end{tikzpicture}
  \caption{Sketch of the simulation box of one cell with the energetically more attractive region in yellow
      and the central region of size $L$ in white. Periodic boundary conditions
      are applied along the cell boundaries in $x$ and $y$-direction.}
  \label{fig:sim_box}
\end{figure}

One cell with an energetically more attractive environment is used as the simulation box with periodic boundary conditions
in the substrate plane ($x$- and $y$- direction) as presented in \figref{\ref{fig:sim_box}}.
\begin{figure}[t]
  \centering
  \begin{tikzpicture}[ line width=1pt]
  \draw[->] (0,0) -- (4,0) node[anchor=north] {time};
  \draw[->] (0,0) -- (0,4);
  \node[rotate=90, above=0.1cm] at (0,1.5) {{ coverage}};
  \draw[dotted] (3,0) node[below] {$t_{\mathrm{end}}$} -- (3,3.2);
  \draw[dotted] (0,3) node[left] {$2$ ML} -- (3.2,3);
  \draw[thick] (0,0) -- (3,3);
  \draw[thick] (1,1) -- (2,1) node[right] {$F$} -- (2,2);
\end{tikzpicture}
  \caption{Simulation time scale is directly connected to the total coverage.}
  \label{fig:cov_time_schema}
\end{figure}
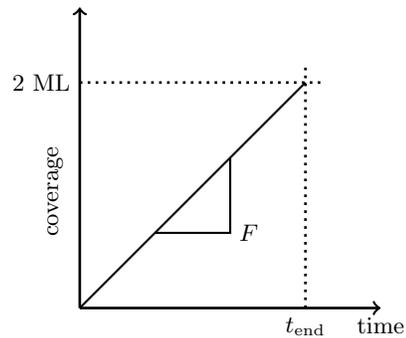
Simulations are done for different cell sizes $L$ and a fixed pre-pattern stripe width of $2\cdot 10a$.
The substrate is composed  of one layer including pre-pattern and substrate sites, which are fixed during the simulation.
The diffusion activation energy on the plane and homogeneous substrate is renormalized, in order to improve
the simulation speed \cite{Strobel2001domain,Mueller2002template}, by accepting every diffusion step of an ad-atom
on the substrate. The simulation starts with a clean substrate. During one
Monte Carlo (MC) step, corresponding to the time step $\Delta t$, every particle on the substrate
attempts one 3D nearest-neighbor move to a new position. The jump to the new position is accepted
according to the standard Metropolis criterion \cite{Metropolis53}.
Despite the simplicity of the model, it provides accurate dynamics as shown by an equivalent model (except zero flux and
desorption is not suppressed) for wetting film instabilities on patterned substrates \cite{Tewes2017comparing}.
Furthermore, all moves are discarded to already occupied sites or after which a particle has zero energy.
In this way possible desorption processes are suppressed. After finishing the MC step, $n$ particles are
positioned on the substrate. Here, $n \in \mathbb{N}$ is a Poissonian random number from a distribution with a mean
value of $\bar{n} \in \mathbb{R}_+$, which is related to the average flux \textit{via}
\begin{equation}
  \label{eq:defflx}
  F = \frac{ \bar{n}}{(A \Delta t)}.
\end{equation}
With $A = (L+20a)^2$ defined as the surface area. Therefore, the flux can be varied continuously. The added
particles are directly attached on a randomly chosen free location of the surface. The resulting time-dependence
of the coverage (number of particles over $A$) is sketched in \figref{\ref{fig:cov_time_schema}}.

All simulations are repeated 2000 times (representing a set of 2000 different cells) in order to obtain
a good statistical description. This large number also acknowledges that the nucleation process
has a significant random component which has to be appropriately averaged out for the analysis.


\section{Theory of single-cluster formation}
\label{sec:theo_sing_cluster}

In \secref{\ref{sec:flux_dep_rho}} a scaling relation is derived for the single-cluster formation with
the feature of one cluster per cell $N=1$, like it is presented in \figref{\ref{fig:exp_fig}} left.
Since the cluster creation process involves stochastic contributions,
nucleation control is defined by the condition that on average one cluster \textit{per} cell is present,
i.e. $\avN = 1$ after the deposition of 2 ML.
The scaling relations, to be derived below, are based on the insight about nucleation control, described
in our previous work  \cite{paperZero}. Therefore, these results are briefly summarized in
\secref{\ref{sec:crit_dens}} first.

\subsection{\label{sec:crit_dens} Existence of a critical density}

Assuming the particle density distribution $\rho(x,y,t)$, with $x,y \in [0,L]$ under the condition that no nucleation
has occurred until time $t$, i.e. no increase of the local density due to nucleation occurs,
then $\rho(x,y,t)$ corresponds in a good approximation to the ad-atom density.
Thus, the mean-field approach based on Burton-Cabrera-Frank theory can be applied for the case of complete
condensation \cite{Burton1951299,Vvedensky199144, Kalischewski200878}. Applying adsorbing boundary
conditions, cf. \secref{\ref{sec:sim_setput}}, the stationary solution
\begin{equation}
\label{eq:stat_sol}
\rho \lowt{stat} (x,y) = \frac{F L^2}{D}  g\left (\frac{x}{L} , \frac{y}{L}  \right )
\end{equation}
is derived, with the explicit form of function $g(x/L,y/L)$ found in \cite{paperZero}.
The stationary density $\rho \lowt{stat} (x,y)$ displays a significant maximum in the center of the cell.
The density $\rho^*\equiv \rho \lowt{stat} (x=L/2,y=L/2)$ at this maximum is
\begin{equation}
  \label{eq:num_sum}
\rho^* =  \frac{F L^2}{D}~C,
\end{equation}
with the constant $C = g(\frac{1}{2}, \frac{1}{2})$.
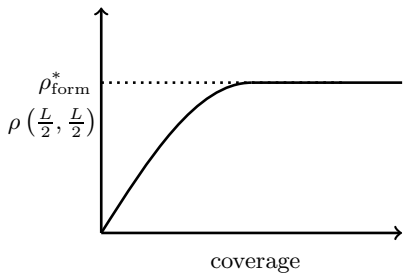
\begin{figure}[t]
  \centering
  \begin{tikzpicture}[ line width=1pt]
  \node[below=0.1cm] at (2,-0.1) {{ coverage}};
  \draw[->] (0,0) -- (4,0);
  \draw[->] (0,0) -- (0,3);
  \node[left=0.2cm, above=0.3cm] at (-0.5,0.85) {{ $\rho \left ( \frac{L}{2},\frac{L}{2} \right )$}};
  \draw (0,0) sin (2,2);
  \draw (2,2) -- (4,2);
  \draw[dotted] (0,2) node[left] {$\rho^*\lowt{form}$} -- (3.2,2);
\end{tikzpicture}
  \caption{\label{fig:rho_form_schema} Evolution of $\rho(x=L/2,y=L/2,t)$ with coverage in the \emph{cluster-free
    sub-ensemble} for a parameter set with $\avN=1$. The long-time limit of $ \rho(L/2,L/2,t)$,
    denoted $\rho^*$, defines the value of $\rho^*\lowt{form}$.}
\end{figure}
For the interesting parameter set expecting $\avN = 1$ a sketch of the time-dependence $\rho(x=L/2,y=L/2,t)$ is
shown in \figref{\ref{fig:rho_form_schema}}.
Note, that in simulations only those realizations (cells) contribute to the density calculation for which
until time $t$ no nucleation process has occurred (\emph{cluster-free sub-ensemble}).
The long-time limit of $ \rho(L/2,L/2)$ in the simulations agrees very well with the analytically calculated value
of $\rho^*$; cf. \eqaref{\ref{eq:num_sum}}. We denote this value $\rho^*\lowt{form} = \rho^* = \rho(L/2,L/2) $,
which is obtained for this special parameter combination with $\avN = 1$.
This value is of major importance when analyzing simulations with a different value of $L$ and a fixed $D/F$.
For smaller $L$, implying $\rho^* < \rho^*\lowt{form}$ cluster formation is strongly suppressed, i.e. $\avN < 1$.
In contrast, in the opposite limit many clusters start to grow as soon as $\rho(x=L/2,y=L/2,t)$ approaches
$\rho^*\lowt{form}$. Thus, the stationary regime is never reached because of preceding nucleation events.
As a consequence $\rho^*\lowt{form}$ can be interpreted as the critical density where nucleation sets in. Since the
nucleation rate scales with a high power of the density (depending on the size of the critical nucleus),
this observation is in agreement with mean-field nucleation theory \cite[chap. 2]{islands_mounds_atoms}.

As a consequence of this interpretation, it comes out that for $\avN = 1$ the nucleation process
occurs close to the center of the cell; see \figref{\ref{fig:exp_fig}}, since only for a small spatial regime
one has $\rho(x=L/2,y=L/2,t) \approx \rho^*\lowt{form}$.

To obtain the density $\rho(x,y,t)$ from the simulation results the projection is used of the three-dimensional deposited
particle distribution onto the surface plane $P(i_x,i_y)$. If the position $(i_x,i_y)$
is occupied by a deposited particle we choose $P(i_x,i_y) = 1 $, otherwise $P(i_x,i_y)=0$. The
density $\rho (i_x,i_y)$ is then defined as the ensemble average of $P(i_x,i_y)$, i.e.
\begin{equation}
  \label{eq:sim_dens}
  \rho (i_x,i_y) = \avr{ P(i_x,i_y) }.
\end{equation}

\subsection{Flux-dependence of the critical density}
\label{sec:flux_dep_rho}
On a theoretical basis one can estimate the flux-dependence of the critical density $\rho^*\lowt{form}$.
According to mean-field nucleation theory
the local nucleation rate scales like  $\Gamma \lowt{loc} \propto \rho^{I+1}$ if $I$ denotes the critical nucleus size.
Note that in \eqaref{\ref{eq:stat_sol}}
the spatial dependence is given by $x/L$ and $y/L$, respectively. As a consequence, the spatial range close to the center
the maximum of the density distribution,
where cluster formation is most likely, i.e. $\rho \lowt{stat} (x,y) \approx \rho^*\lowt{form}$, is proportional
to the total area $L^2$. Thus, for the interesting parameter set with $\avN = 1$ one obtains for the total nucleation rate
\begin{equation}
  \label{eq:gammatot}
  \Gamma \lowt{tot} \propto \rho^{* I+1}\lowt{form}(F) L^2.
\end{equation}
The possible flux-dependence is explicitly indicated.
This relation only holds under the condition that no nucleation has occurred so far.

The nucleation rate in this case ($\avN = 1$) does not depend on the flux, as will be shown in the next
\secref{\ref{sec:quanlit_behav}}. As an immediate consequence one has the simple scaling $\Gamma_{tot} \propto F$,
i.e. when doubling the flux and appropriately modifying the length $L$ to keep $\avN = 1$ also the total nucleation
rate will become twice as large. From \eqaref{\ref{eq:num_sum}} it follows that there is a direct relation between
the chosen length scale $L$ and the flux $F$. Choosing $L^2 \propto \rho^*\lowt{form}(F)/F$ one explicitly keeps
track of the flux dependence in order to keep $\rho^* \equiv \rho \lowt{stat}$ constant. Inserting both relations
into \eqaref{\ref{eq:gammatot}} one gets $F \propto \rho^{*I+2}\lowt{form}(F)/F$ and thus
\begin{equation}
  \label{eq:rhof}
  \rho^*\lowt{form}(F) \propto F^p
\end{equation}
with $p = 2/(I+2)$
as an immediate consequence one can predict how one has to vary the grid size $L$ in order to keep $\avN = 1$ when varying the flux.
In general \eqaref{\ref{eq:num_sum}} yields $L^2 F \propto \rho^*\lowt{form}(F)$. Together with \eqaref{\ref{eq:rhof}} this yields
\begin{equation}
  \label{eq:LpropF}
  L \propto F^{-q}
\end{equation}
with
\begin{equation}
  q = \frac{1-p}{2} = \frac{I}{2(I+2)}.
\end{equation}
As a matter of fact we end up with the same scaling expression given by Ranguelov {\em et al.} \cite{Ranguelov200775}, who used
a related one-dimensional approach to analyse the island creation on stepped surfaces. The resulting scaling exponent of
$q = I/(2(I+2))$ is in accordance with island density scaling, one would expect on a homogeneous substrate
for complete condensation \cite{Hanbuecken1984399}.

\section{Growth of cluster - qualitative behavior}\label{sec:quanlit_behav}

The observable $\avN$ is choosen to characterize the structures on substrate with adsorbing
grid pre-patterns.
For very small values of $L$, in comparison to $L$ with $\avN = 1$ and fixed $D/F$, most
particles are adsorbed at the pre-pattern boundaries and no cluster are formed in the center of the cell.
In contrast, for very large $L$ the boundary only plays a minor role and mean-field nucleation behavior
can be observed \cite[chap. 2]{islands_mounds_atoms}. Of particular interest is the intermediate value of $L$  for
which $\avN$ is unity, i.e on average a single cluster is formed \textit{per} cell. In practice the
chosen flux is tuned for a fixed $L$, until $\avN \approx 1$.

\begin{figure}[tb]
  \includegraphics[width=0.48\textwidth]{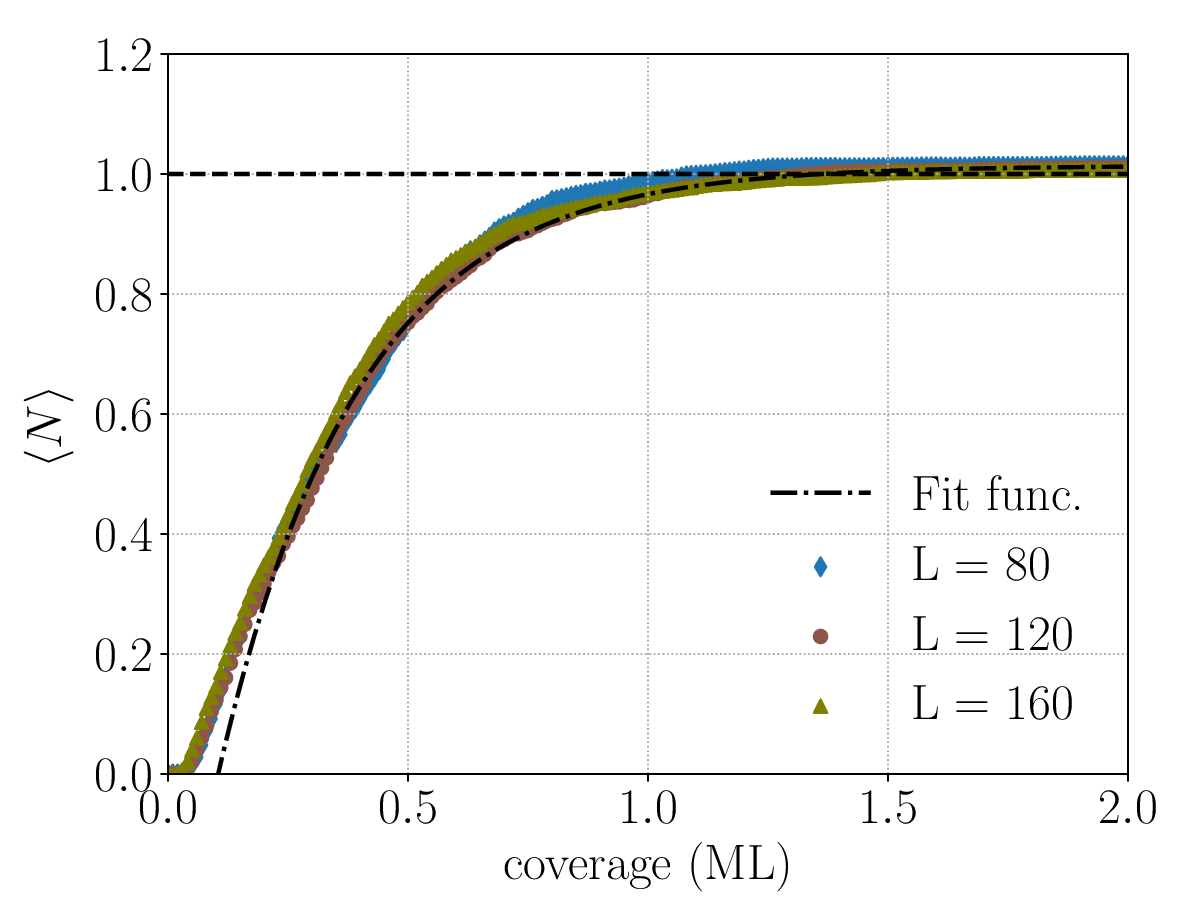}
 \includegraphics[width=0.48\textwidth]{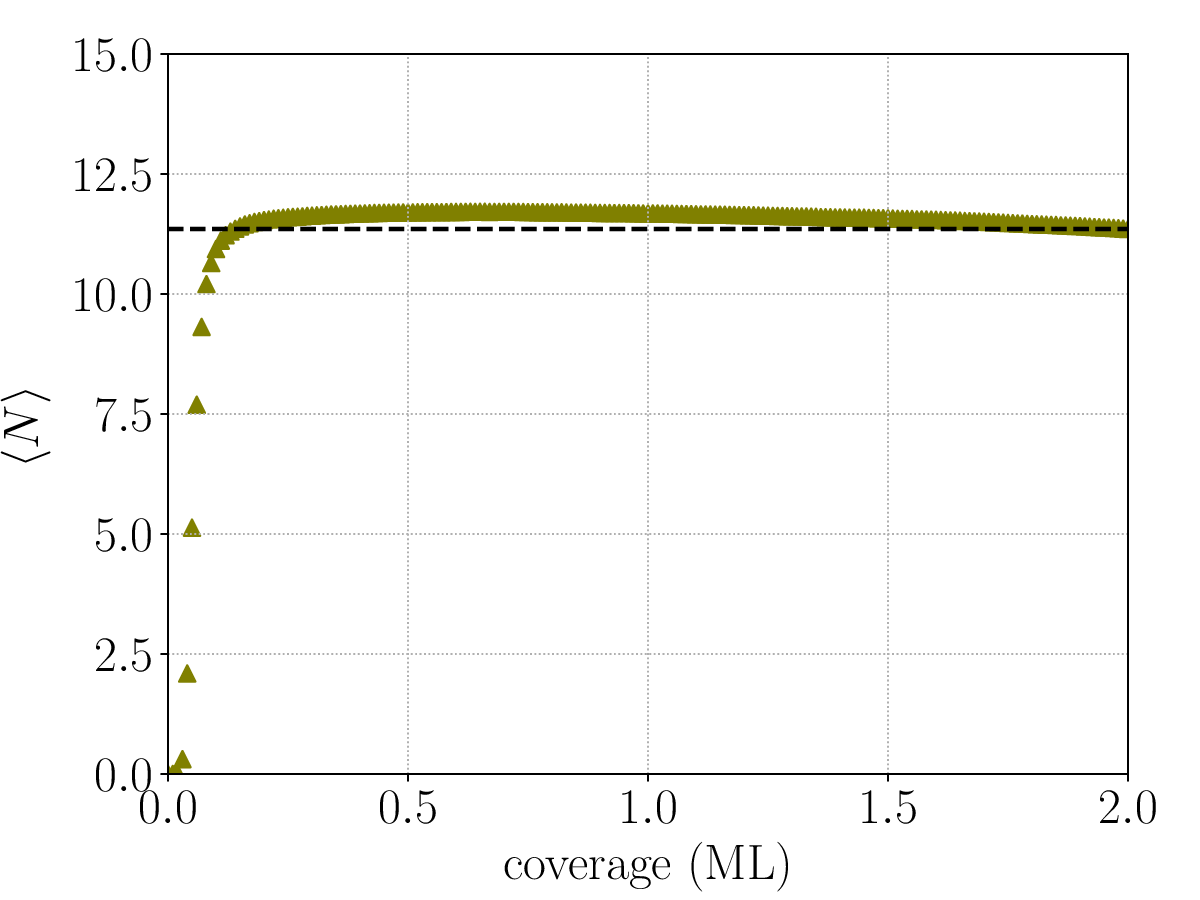}
  \caption{\label{fig:figNvsML_l100} Average number of clusters in dependence of the coverage.
 {\bf Top}: Different $L$ values in the case where $\avN$ approaches one. The respective values
 for the corresponding flux are listed in \tabref{\ref{tab:nOne_data}}. For $L=120$ an
 exponential fit of the form $f(x) = c_0 - c_1\exp{(-c_2x)}$ to the long-time behavior
 is included. The dashed line marks $\avN = 1$.
 {\bf Bottom:} $\avN $ on grid lattice size of $L= 160$ and flux of $F=5.85 \times 10^{-6}/(a^2 \Delta t)$.
 The dashed line marks $\avN = 11.4 $ reached after the deposition of 2.0 ML.}
\end{figure}

\begin{table}[h]
  \caption{\label{tab:nOne_data} Data for $\avN \approx 1$ with cell size $L$, the corresponding flux $F$,
    the average cluster number \textit{per} cell $\avN$ (sum of all clusters found in 2000 cells divided by the
    number of cells, here 2000), the standard deviation $\sigma$ of the distribution of clusters per
    cell and yield (fraction of cells with exactly one cluster).}
 \begin{ruledtabular}
  \begin{tabular}{ccccc}
   $L/a$ &  $F$ in $10^{-6}/(a^2 \Delta t)$ & $\avN$ & $\sigma$ & yield in \% \\ \hline
    40  & 36.11 & 1.026 & 0.208 & 95.6  \\
    60 &  12.03 & 1.014  & 0.215  & 95.4  \\
    80 &  5.85 & 1.004 & 0.212 & 95.7  \\
    120 & 2.14 &  1.011 &  0.219 & 95.2  \\
    160 & 1.08 & 1.006 & 0.212 & 95.5  \\
  \end{tabular}
 \end{ruledtabular}
\end{table}

The value of $\avN$ depends on the coverage of the surface. In \figref{\ref{fig:figNvsML_l100}} (top)
the diagram $\avN$ against the coverage in monolayer (ML number of particles
\textit{per} full surface coverage) is shown for different values of $L$. The respective values for the
corresponding flux are listed in \tabref{\ref{tab:nOne_data}}.
Interestingly, when expressing the time in terms of the coverage, one observes no
major dependence on the cell size or the flux,
respectively. As a consequence, the doubling of the flux gives rise to a doubling of the nucleation rate under the condition
of an appropriate choice of the new $L$ to guarantee $\avN = 1$.

The exponential fit, also shown in \figref{\ref{fig:figNvsML_l100}} (top), works well for a coverage larger than
0.2 ML. In this regime  the nucleation rate is independent of time. Note that in the long-time regime for
a coverage over 1.5 ML the value of $\avN$ reaches a plateau, i.e. almost no new clusters are formed. Of course,
in this regime the existing individual clusters are still growing. As a consequence, the value of $\avN$, determined
for a coverage of 2.0 ML, is very well defined and insensitive on the chosen coverage.  The exponential fit does not
work for low coverage. A detailed analysis of the short-time behavior, however, is beyond the scope of this work.

In contrast, for  $F=5.85 \times 10^{-6}/(a^2 \Delta t)$ and  $L= 160$  on average more than 11 clusters are
created \textit{per} cell; see \figref{\ref{fig:figNvsML_l100}} (bottom). Almost all nucleation proceeds
before 0.1 ML are deposited. The curve displays a maximum at 0.4 ML and slightly decreases for longer times,
indicating some coarsening effect which, however is small. Therefore  $\avN$ after the deposition of 2.0 ML is
chosen to characterize the formation.

\section{Nucleation and position control}
\label{sec:nuc_and_pos_control}

\begin{figure}[t]
  \includegraphics[width=0.5\textwidth]{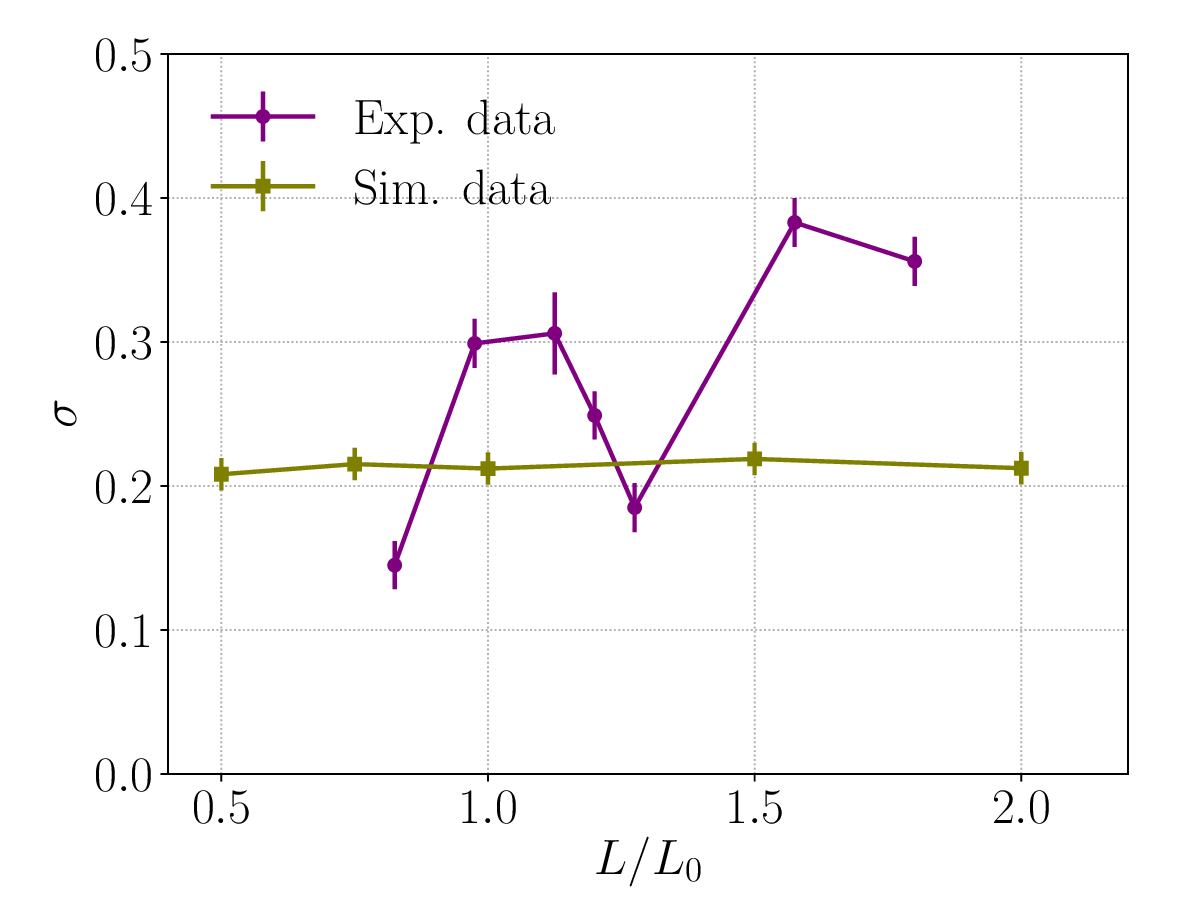}
  \caption{\label{fig:figN1std} Standard deviation of $\avN $ from experiment and simulation, for simulation data $L_0 = 80a$
    was used and for the experimental data $L_0 =\SI{8/3}{\um}$. The simulation data is also listed in \tabref{\ref{tab:nOne_data}}.
  The errorbars are estimated with $1/(2\sqrt{n})$, with $n$ as the size of the data set.}
\end{figure}

One goal is to generate a structure where ideally in every cell there is exactly one cluster which, furthermore, is
located directly in the center of the cell. In this scenario one might speak of ideal nucleation and
position control, respectively. This implies the choice $\avN \approx 1$. The opposite implication is not correct.
A system with $\avN \approx 1$ may fail to show reasonable nucleation control, e.g. by having many cells with zero and two nuclei.
In the following, the nucleation and position control from the simulations are compared with those from the experiments.
First, to check the quality of nucleation, the standard deviation of the distribution of nuclei per cell is determined. The result
is shown in \figref{\ref{fig:figN1std}}.
It turns out that the quality of the nucleation control from the simulations is independent of the chosen $(F,L)$-pair.
A standard deviation of 0.2 together with an average value of unity ($\avN \approx 1$) means that in approx. 95\% of all cells there is
a single cluster, as listed in \tabref{\ref{tab:nOne_data}}, whereas in the remaining cells there is the same number
of cells with no or two clusters. The independence from $L$  just reflects the scaling properties, discussed above.
Thus, after rescaling the time- and length scale the nucleation behavior is basically identical.

It is very promising that the standard deviations, seen experimentally, are very close to the simulated ones. This
shows that the lattice model represents the key properties of the nucleation behavior, including that of the actual molecules
in the experiment.
The fluctuations as a function of $L$ are much larger than expected from statistical reasons.
Experimentally, reasons for fluctuation of experimental data could be related to a slightly uncontrolled coverage
and contamination on the substrate.

\begin{figure}[t]
\includegraphics[width=0.5\textwidth]{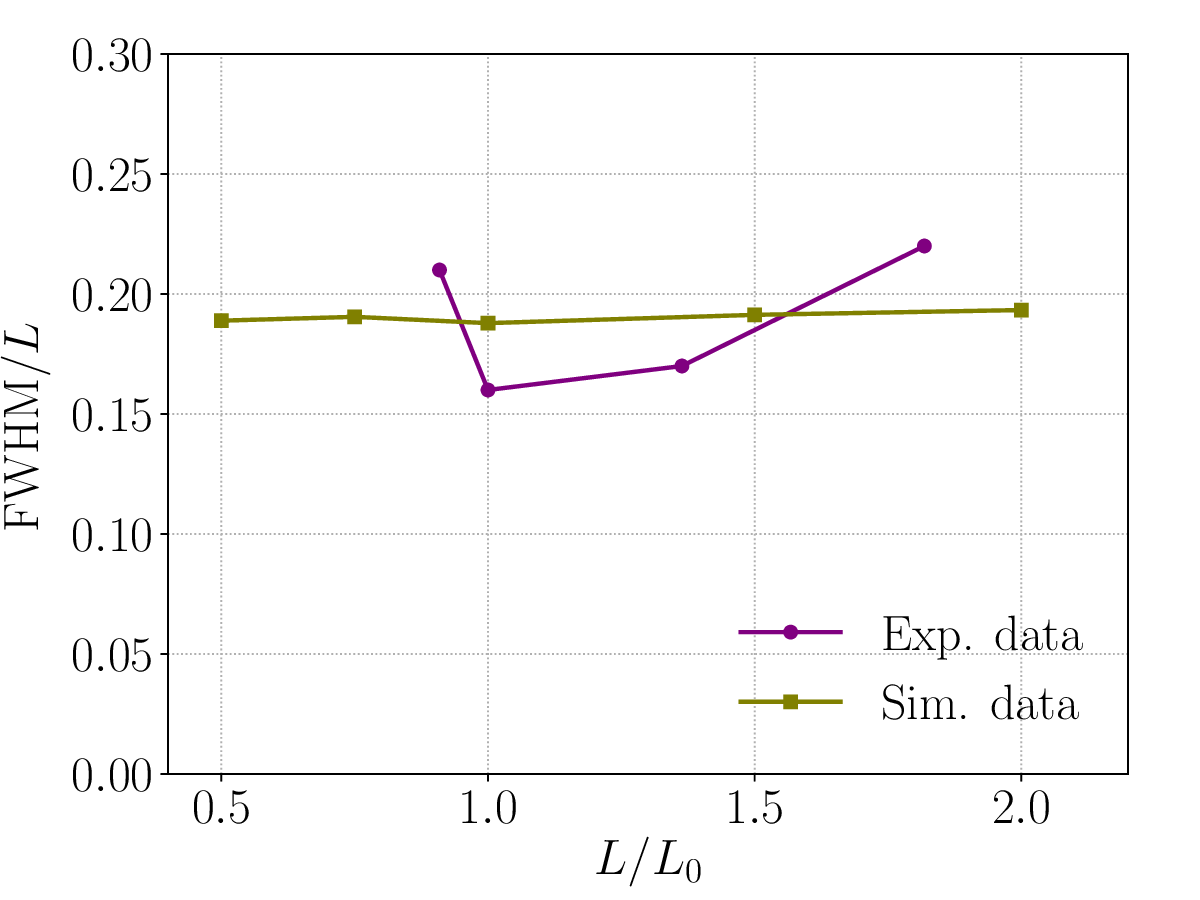}
\caption{\label{fig:FWHMvsL}Full width at half maximum of the center of mass distribution
  corresponding to the center of the cell $\left(L/2,L/2 \right)$ scaled by the cell size $L$ versus $L$.
  The scaling factor $L_0$ is the same as in \figref{\ref{fig:figN1std}}}
\end{figure}

Second, to analyse the position control we identify for each cluster the center of mass and then analyse the spatial
distribution of these centers. This distribution has its maximum in the middle of the cell.  In the next
step we determine the full width at half maximum (FWHM). Due to the scaling arguments, already discussed above, we
expect that this width should scale with $L$. As shown in \figref{\ref{fig:FWHMvsL}}, the ratio FWHM/$L$ is
independent of $L$. Again, simulations and experiments basically displays the same values. Furthermore this
ratio is small which is a quantitative confirmation that not only nucleation control but also position control
works very well.

At last, we looked at the average cluster sizes $\langle s \rangle$ and the relative width of the cluster size
distributions $\omega_s$ in two and three dimensions, see \tabref{\ref{tab:csize_dist}}.
By consideration of the average 2D (size of 2D projection, see \secref{\ref{sec:crit_dens}}) and
3D (number of particles) sizes, structural changes can be identiﬁed. These changes are expected in
kinetically driven processes. In general, one would expect a scaling with $L^2$. Interestingly, the data in
\tabref{\ref{tab:csize_dist}} shows a weaker increase. The 3D size scales with $L^{2.00-0.18}$ and the 2D size
with $L^{2.00-0.68}$. The strong deviations for the 2D size from a quadratic scaling can be qualitatively
related to the observation that the form of the cluster changes with the external flux. For high flux (small $L$)
the cluster is relatively flat whereas for low flux (large $L$)  it starts to become more compact, displaying a
more spherical shape. For lower flux the system has more time to approach the free energy minimum.
Interestingly, the relative width of the size distributions in 3D and 2D is almost $L$ independent.

\begin{table}[h]
  \caption{\label{tab:csize_dist} Average cluster size in 3D (number of particles), 2D
                                  (size of 2D projection, see \secref{\ref{sec:crit_dens}})
                                  and the relative widths $\omega_s$ of the cluster
                                  size distributions  for $\avN \approx 1$ on different grid sizes $L$.}
 \begin{ruledtabular}
  \begin{tabular}{cccccc}
     $L/a$                     & 40   & 60     &  80    & 120   &  160   \\ \hline
      3D: $\langle s \rangle$   & 835  & 1655   &  2844  &  5795  &  9917 \\
      3D:         $\omega_{s}$  & 0.24 & 0.26   & 0.23   &  0.24 &  0.24 \\ \hline
      2D: $\langle s \rangle$   & 207  &  343   & 514    &  862  &  1276 \\
      2D:         $\omega_{s}$  & 0.20 & 0.22   & 0.20   &  0.22 &  0.21 \\
  \end{tabular}
 \end{ruledtabular}
\end{table}

\section{Flux dependence}
\label{sec:flux_dep}
\begin{figure}[t]
\includegraphics[scale=0.45]{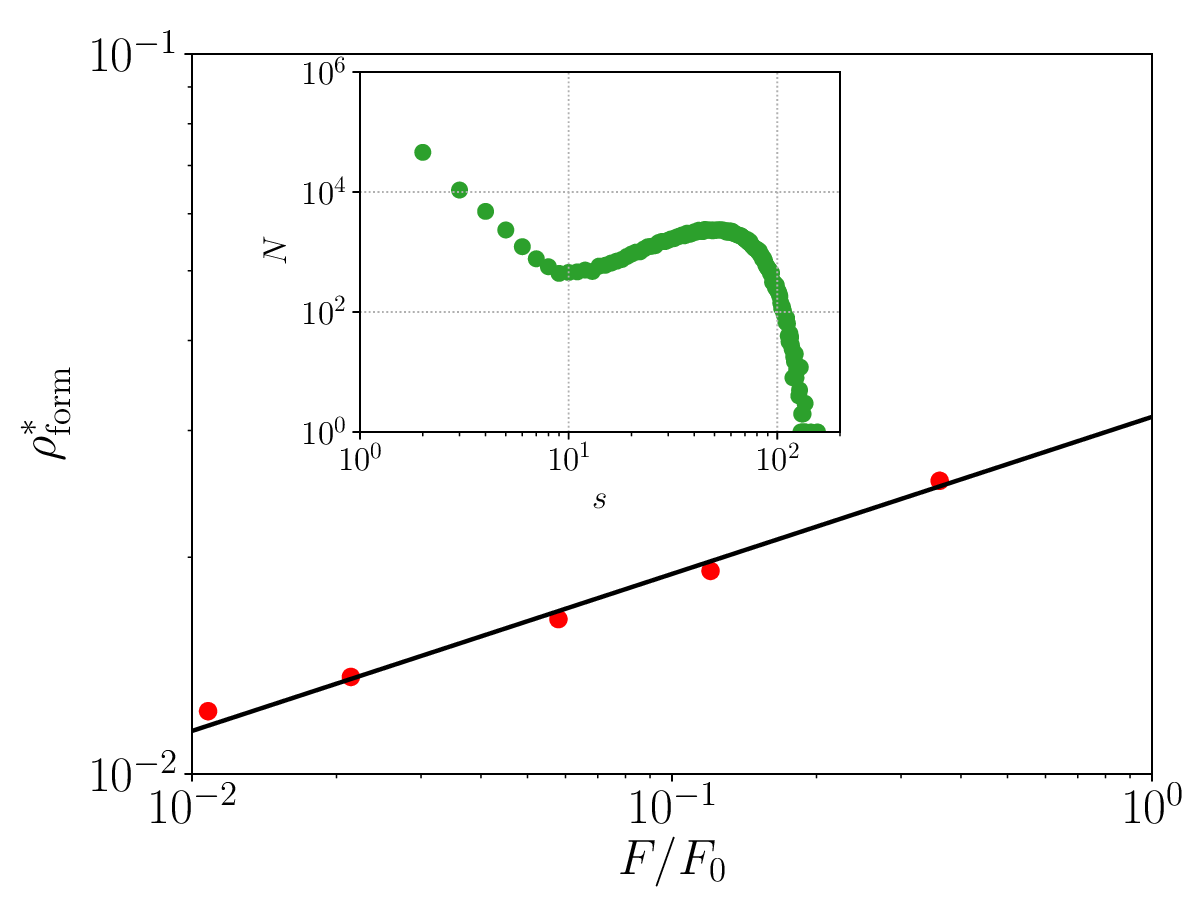}
\caption{\label{fig:figN1L} Critical density $\rho^*\lowt{form}$ against $F/F_0$ ($F_0 = 10^4 \frac{1}{a^2 \Delta t}$)
  for appropriately chosen cell sizes, corresponding to $\avN \approx 1$. Included is a power law fit with an exponent of $p=0.22$,
  cf. \eqaref{(\ref{eq:rhof})}.
  In the inset the double-logarithmic histogram of the 3D cluster sizes $s$ is shown which is observed from simulations
  on un-patterned substrates. The minimum is found for a size of $s=9$. The analysis is done
  for $F=12.03 \frac{1}{a^2 \Delta t}$ and a coverage of 0.1 ML.}
\end{figure}

As discussed in \secref{\ref{sec:flux_dep_rho}} the key observable to characterize the underlying flux dependence of the cluster
formation is the critical density $\rho^*\lowt{form}$. Its values have been determined for different $(F,L)$-pairs,
corresponding to $\avN \approx 1$ as given in \tabref{\ref{tab:nOne_data}}.
For this purpose the stationary density $\rho^*$ is determined under the condition $\avN \approx 1$,
like it is described in \secref{\ref{sec:crit_dens}}.
The extracted $\rho^* \lowt{form}$ values are shown in \figref{\ref{fig:figN1L}}.

As expected from \eqaref{\ref{eq:rhof}}, a power-law relation is observed between $\rho^* \lowt{form}$ and $F$ with an
exponent of $p=0.22$. In \eqaref{\ref{eq:rhof}} this exponent is expressed in terms of the critical cluster size $I$.
The value of $I$ is estimated by analyzing the distribution of cluster sizes for a simulation on an
homogeneous (un-patterned) substrate. The minimum can be taken as a measure for $I$ \cite{Amar199574}.
In this way $I=9$ is obtained. According to \eqaref{\ref{eq:rhof}} this would give rise to an exponent
of 0.18 which is close to the observed value of 0.22.

\begin{figure}[t]
   \centering
   \includegraphics[width=0.45\textwidth]{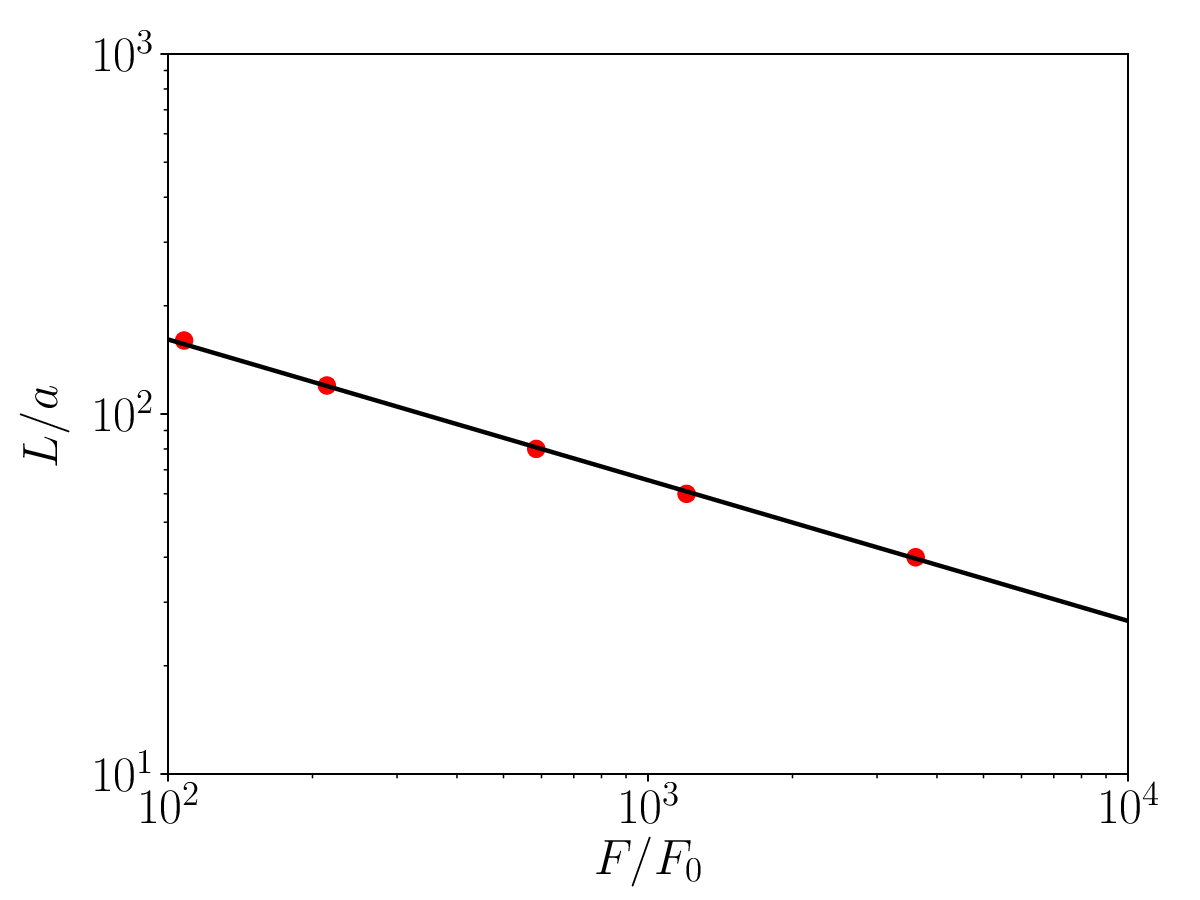}
   \includegraphics[width=0.45\textwidth]{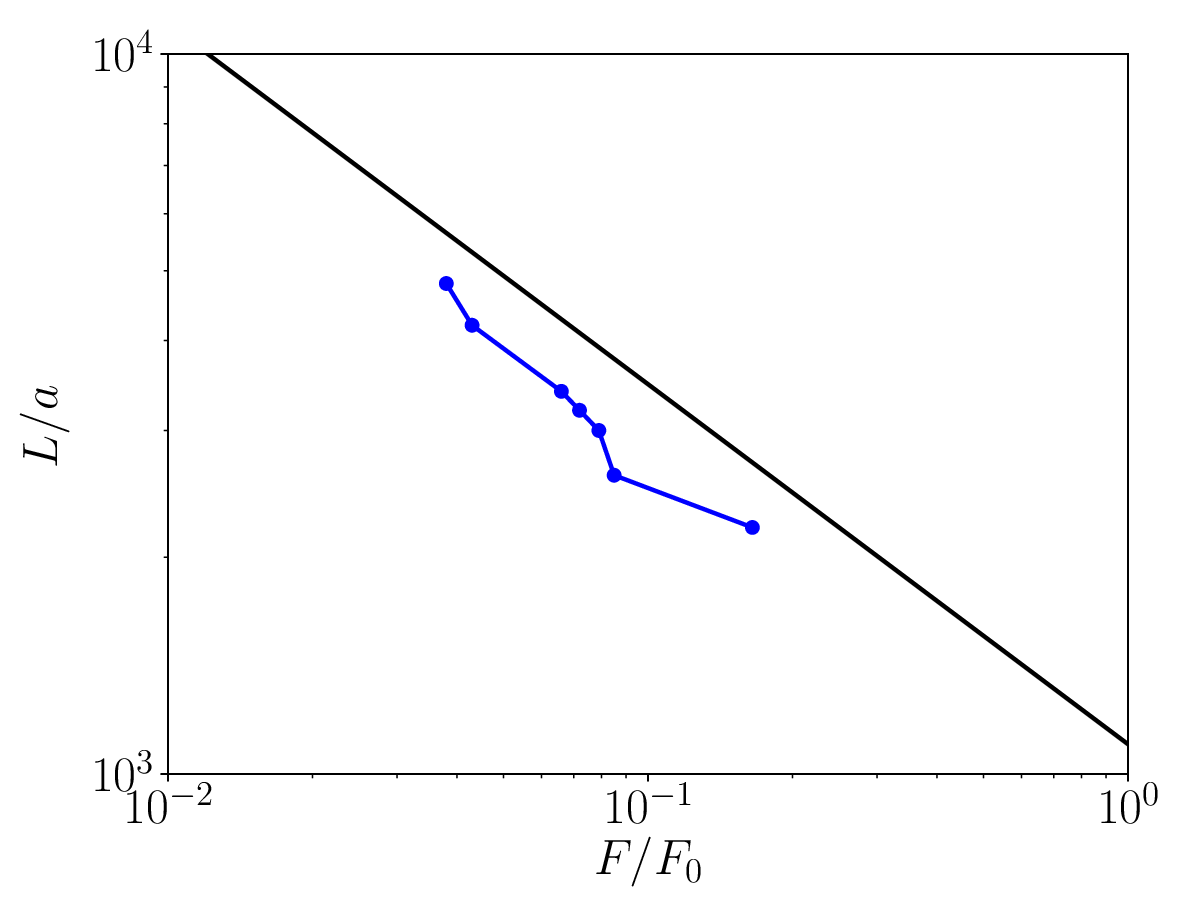}
   \caption{\label{fig:LvsF} Double-logarithmic representation of  $L/a$  against $F/F_0$  for $\avN = 1$.
     {\bf Top:} Simulation data, included is a power law graph with an exponent of $q=0.39$ and $F_0 = 10^8 /(a^2 \Delta t)$.
     {\bf Bottom:} Experimental data with $F_0=\SI{1.0}{\nm\per\minute}$.
     In this case the value of $L$ is expressed in units of the elementary length scale of the organic molecule which
     it approximated as $a=\SI{1}{\nm}$. Included is a power-low with exponent $q=0.5$.}
\end{figure}

In the next step the dependence of the system size on the flux is predicted. As discussed in Sect \ref{sec:flux_dep_rho} the
slope should be given by $q=(1-p)/2 = 0.39$. The simulated data can be described very well by this exponent, see
\figref{\ref{fig:LvsF}} top. Using $I = 9$, one would end up with an exponent of 0.41, which is also very close to observed slope.

In the similar plot for the experimental {$L(F)$-dependence}, presented in \figref{\ref{fig:LvsF}} bottom,
an exponent of $q \approx 0.5 $ is obtained under the assumption of the scaling as in \eqaref{\ref{eq:LpropF}}, which
is only valid for the complete condensation regime. In this case, $q \approx 0.5 $ corresponds to  a very large critical
cluster size $I$. But, due to the fact of unknown condensation mechanism in the experiment, which determines the scaling
of $L(F)$ \cite{Hanbuecken1984399}, we cannot determine the exact critical cluster size. Nonetheless this shows that
the scaling in general is not limited to a specific regime of condensation.

\section{Scaling beyond $\avN = 1$}\label{sec:nuc_to_std_nuc}
\subsection{{$F/D$-scaling}}

It is known from mean-field nucleation theory that the number of clusters in the stationary long time limit on a plain substrate
is a function of $F/D$ \cite[chap. 2]{islands_mounds_atoms}.
In the present case this would correspond to the limit of large $\avN$ where the influence of the boundaries hardly matter.
From our theoretical approach to nucleation control (\eqaref{\ref{eq:num_sum}}) we also expect a perfect $F/D$ scaling
for $\avN=1$.

\begin{figure}[t]
  \includegraphics[width=0.48\textwidth]{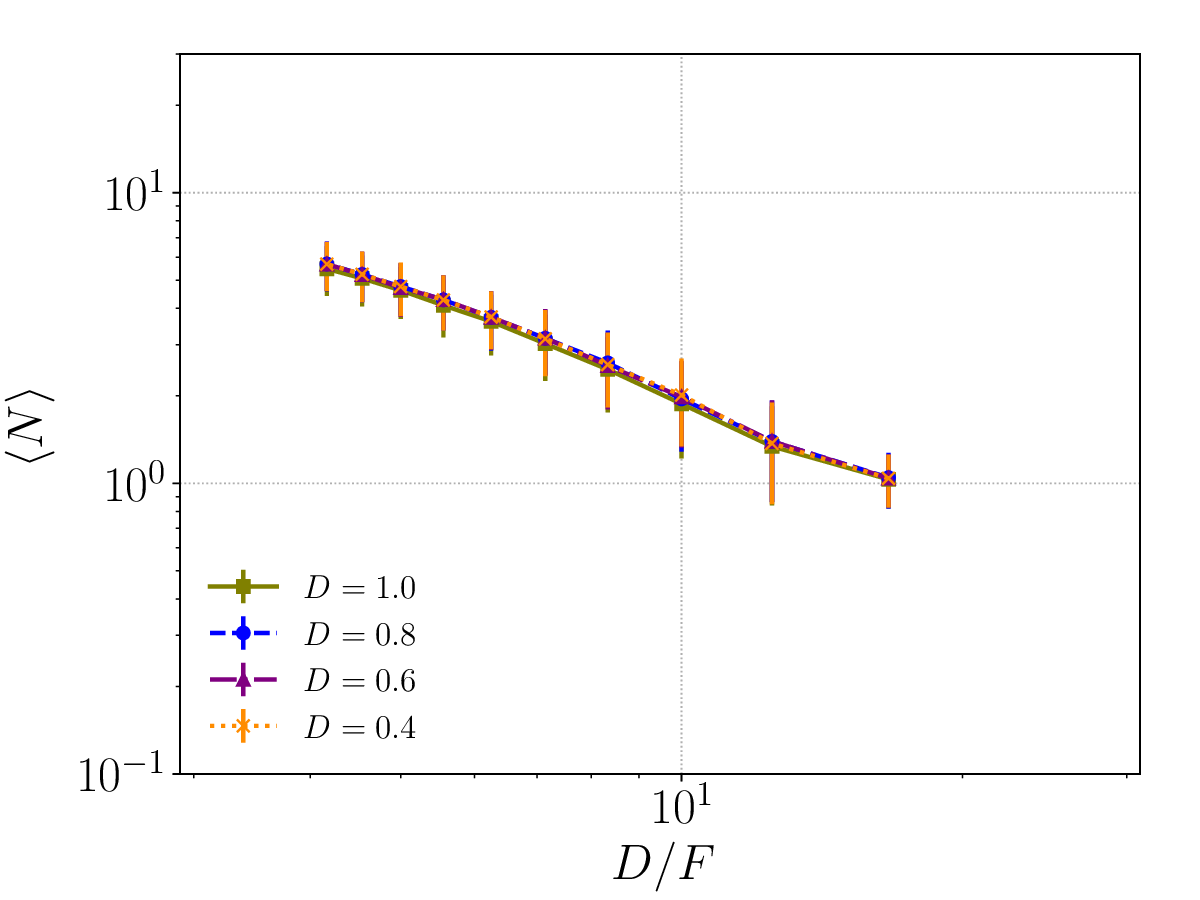}
  \caption{\label{fig:figNvsDoF} Average number of clusters versus $D/F$ for different values of the diffusion
    parameter as indicated in the figure and the corresponding flux $F$ on a cell of size $L=80a$ in a double-logarithmic
    representation. The flux is scaled by factor $10^{4} a^2 \Delta t$.}
\end{figure}

Here, this scaling with $F/D$ is verified whether it holds for the large range of $\avN$-values.
For this purpose the values of $D$ and $F$  are varied individually.
Specifically, a diffusion parameter $D$ is introduced as the probability to make a MC move. Therefore
$D \in [0,1]$ for $D=1.0$ the KMC dynamic is as before, for $D=0.5$ on average every second try for a MC step is denied, for $D=0.0$
the system does not move at all. Thus, for decreasing $D$ the dynamics in the system is slowed down.
In \figref{\ref{fig:figNvsDoF}} the average number of stable clusters $\avN$ is displayed against $D/F$.
Note that a perfect scaling is observed for all values of $\avN$.
As a practical consequence, $D=1.0$ can always be chosen in order to optimize the efficiency of the MC simulations
but nevertheless it covers all possible diffusion constants.

\subsection{Impact of length $L$}

\begin{figure}[t]
  \includegraphics[scale=0.48]{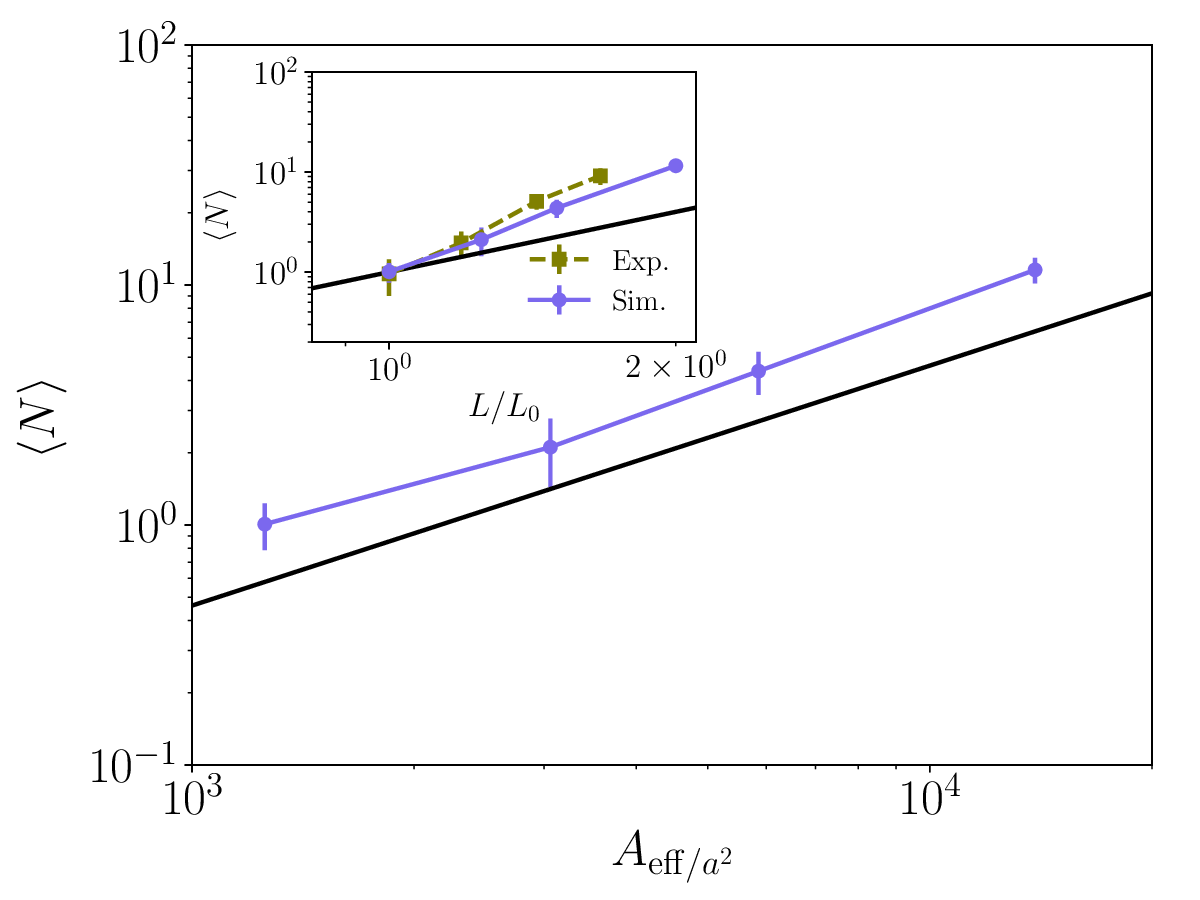}
  \caption{\label{fig:Nscale} Cluster number scaling against the effective nucleation area for a fixed
    flux $F=5.85 \times 10^{-6}/(a^2 \Delta t)$. The solid line is marking
    a power law with a slope of one.
    In the inset the number of clusters is plotted as function of $L/L_0$ ($L_0 = \SI{4.2}{\micro \meter}$ or $L_0=80a$)
    in the double-logarithmic representation. The error bars represent the standard deviation. The solid line represents
    a power law with a slope of two.}
\end{figure}

Finally, the scaling of $\avN$ is discussed for a fixed flux $F=5.85 \times 10^{-6}/(a^2 \Delta t)$ and
different cell sizes $L$ following the experimental results in \figref{\ref{fig:exp_fig}}.
The data is shown in \figref{\ref{fig:Nscale}}.
Is it possible to judge from which system size the boundary conditions only have a minor impact on
the nucleation behavior? Without pre-pattering one would expect the simple scaling $ \avN \propto L^2$. For
fixed flux values we show the relation between $\avN$ and $L$ in the inset of \figref{\ref{fig:Nscale}}, both
for the simulation and the experimental data. Interestingly, both data sets display a similar slope in the
double logarithmic representation. It is, however, larger than two. These deviations from the expected quadratic scaling
on the un-pattern substrate have their origin in the limited spatial region where nucleation can occur.
This area, denoted as $A\lowt{eff}$, is determined via the condition that the particle density of the overall
ensemble after the deposition of 2 ML is higher than $0.1~1/a^2$.
Plotting $\avN$ as a function of $A\lowt{eff}$ yields a slope of one.
Thus, the number of clusters is extensive and the boundary effect is  reflected
by the presence of a zone where no nucleation can take place.


\section{Conclusions}\label{sec:discussion}

Based on a new experimental way to generate a regular array of clusters, possibly formed by functionalized
organic molecules, we have studied this boundary-induced nucleation control from a theoretical perspective.
In particular the experimental data is compared with the outcome from kinetic Monte Carlo simulations.
More specifically, the quality of nucleation and position control, the flux vs.
length-dependence to guarantee the presence of a single cluster \textit{per} cell and the dependence of the number of
clusters on the effective growth area are analyzed. In all cases a very good agreement is obtained between analytical
description, simulation and experiment.

Furthermore, \textit{via} comparison with analytical expressions for the stationary concentration profile in the
\emph{cluster-free sub-ensemble} and employing key results of mean-field nucleation theory, the flux \textit{versus}
length-dependence can be fully understood. Most properties result from
general scaling relations. One key result is the proportionality of the nucleation rate to the external flux
under the condition that $\avN = 1$.
This scaling behavior was the major ingredient to connect the analytical and the
numerical/experimental pieces of information.
Due to the agreement of independent approaches, we suppose this mechanism of indirect nucleation
control can be generalized and applied to a wide variety of growth scenarios and materials for structure formation on
the mesoscopic scale, as long a diffusion process is present on a homogenious substrate with adjustable sinks.
For appropriatly related values of the diffusivity, grid size, critical cluster size and flux
boundary-induced nucleation control can be achieved experimentally.
We would like to stress, that this scenario explicitly occurs in a non-equilibrium setting.

We hope that combined experimental and theoretical analysis of boundary-induced
nucleation control and the observed generality of this approach may inspire more work along this line. Possible
extensions to multiple component deposition are conceivable.

\begin{acknowledgements}
This work was supported through the Transregional Collaborative Research
Centre TRR 61 (projects B1 and B12) by the DFG.
\end{acknowledgements}


%



\begin{thebibliography}{40}%
\makeatletter
\providecommand \@ifxundefined [1]{%
 \@ifx{#1\undefined}
}%
\providecommand \@ifnum [1]{%
 \ifnum #1\expandafter \@firstoftwo
 \else \expandafter \@secondoftwo
 \fi
}%
\providecommand \@ifx [1]{%
 \ifx #1\expandafter \@firstoftwo
 \else \expandafter \@secondoftwo
 \fi
}%
\providecommand \natexlab [1]{#1}%
\providecommand \enquote  [1]{``#1''}%
\providecommand \bibnamefont  [1]{#1}%
\providecommand \bibfnamefont [1]{#1}%
\providecommand \citenamefont [1]{#1}%
\providecommand \href@noop [0]{\@secondoftwo}%
\providecommand \href [0]{\begingroup \@sanitize@url \@href}%
\providecommand \@href[1]{\@@startlink{#1}\@@href}%
\providecommand \@@href[1]{\endgroup#1\@@endlink}%
\providecommand \@sanitize@url [0]{\catcode `\\12\catcode `\$12\catcode
  `\&12\catcode `\#12\catcode `\^12\catcode `\_12\catcode `\%12\relax}%
\providecommand \@@startlink[1]{}%
\providecommand \@@endlink[0]{}%
\providecommand \url  [0]{\begingroup\@sanitize@url \@url }%
\providecommand \@url [1]{\endgroup\@href {#1}{\urlprefix }}%
\providecommand \urlprefix  [0]{URL }%
\providecommand \Eprint [0]{\href }%
\providecommand \doibase [0]{http://dx.doi.org/}%
\providecommand \selectlanguage [0]{\@gobble}%
\providecommand \bibinfo  [0]{\@secondoftwo}%
\providecommand \bibfield  [0]{\@secondoftwo}%
\providecommand \translation [1]{[#1]}%
\providecommand \BibitemOpen [0]{}%
\providecommand \bibitemStop [0]{}%
\providecommand \bibitemNoStop [0]{.\EOS\space}%
\providecommand \EOS [0]{\spacefactor3000\relax}%
\providecommand \BibitemShut  [1]{\csname bibitem#1\endcsname}%
\let\auto@bib@innerbib\@empty
\bibitem [{\citenamefont {Mahan}(2000)}]{Mahan2000physical}%
  \BibitemOpen
  \bibfield  {author} {\bibinfo {author} {\bibfnamefont {J.~E.}\ \bibnamefont
  {Mahan}},\ }\href
  {http://www.ebook.de/de/product/3601999/john_e_mahan_physical_vapor_deposition_of_thin_films.html}
  {\emph {\bibinfo {title} {Physical Vapor Deposition of Thin Films}}}\
  (\bibinfo  {publisher} {JOHN WILEY \& SONS INC},\ \bibinfo {year}
  {2000})\BibitemShut {NoStop}%
\bibitem [{\citenamefont {Mattox}(2010)}]{Mattox2010handbook}%
  \BibitemOpen
  \bibfield  {author} {\bibinfo {author} {\bibfnamefont {D.~M.}\ \bibnamefont
  {Mattox}},\ }\href
  {http://www.ebook.de/de/product/8235878/donald_m_mattox_handbook_of_physical_vapor_deposition_pvd_processing.html}
  {\emph {\bibinfo {title} {Handbook of Phys. V. Dep. (PVD)
  Processing}}}\ (\bibinfo  {publisher} {William Andrew Publishing},\ \bibinfo
  {year} {2010})\BibitemShut {NoStop}%
\bibitem [{\citenamefont {Michely}\ and\ \citenamefont
  {Krug}(2004)}]{islands_mounds_atoms}%
  \BibitemOpen
  \bibfield  {author} {\bibinfo {author} {\bibfnamefont {T.}~\bibnamefont
  {Michely}}\ and\ \bibinfo {author} {\bibfnamefont {J.}~\bibnamefont {Krug}},\
  }\href@noop {} {\emph {\bibinfo {title} {Islands, Mounds and Atoms}}}\
  (\bibinfo  {publisher} {Springer-Verlag Berlin Heidelberg},\ \bibinfo {year}
  {2004})\BibitemShut {NoStop}%
\bibitem [{\citenamefont {Evans}\ \emph {et~al.}(2006)\citenamefont {Evans},
  \citenamefont {Thiel},\ and\ \citenamefont
  {Bartelt}}]{Evans2006morphological}%
  \BibitemOpen
  \bibfield  {author} {\bibinfo {author} {\bibfnamefont {J.}~\bibnamefont
  {Evans}}, \bibinfo {author} {\bibfnamefont {P.}~\bibnamefont {Thiel}}, \ and\
  \bibinfo {author} {\bibfnamefont {M.}~\bibnamefont {Bartelt}},\ }\href
  {\doibase http://dx.doi.org/10.1016/j.surfrep.2005.08.004} {\bibfield
  {journal} {\bibinfo  {journal} {Surf. Sci. Rep.}\ }\textbf {\bibinfo {volume}
  {61}},\ \bibinfo {pages} {1 } (\bibinfo {year} {2006})}\BibitemShut {NoStop}%
\bibitem [{\citenamefont {Niu}\ \emph {et~al.}(2006)\citenamefont {Niu},
  \citenamefont {Vardavas}, \citenamefont {Caflisch},\ and\ \citenamefont
  {Ratsch}}]{Niu2006level}%
  \BibitemOpen
  \bibfield  {author} {\bibinfo {author} {\bibfnamefont {X.}~\bibnamefont
  {Niu}}, \bibinfo {author} {\bibfnamefont {R.}~\bibnamefont {Vardavas}},
  \bibinfo {author} {\bibfnamefont {R.~E.}\ \bibnamefont {Caflisch}}, \ and\
  \bibinfo {author} {\bibfnamefont {C.}~\bibnamefont {Ratsch}},\ }\href
  {\doibase 10.1103/PhysRevB.74.193403} {\bibfield  {journal} {\bibinfo
  {journal} {Phys. Rev. B}\ }\textbf {\bibinfo {volume} {74}},\ \bibinfo
  {pages} {193403} (\bibinfo {year} {2006})}\BibitemShut {NoStop}%
\bibitem [{\citenamefont {Nurminen}\ \emph {et~al.}(2000)\citenamefont
  {Nurminen}, \citenamefont {Kuronen},\ and\ \citenamefont
  {Kaski}}]{Nurminen200063}%
  \BibitemOpen
  \bibfield  {author} {\bibinfo {author} {\bibfnamefont {L.}~\bibnamefont
  {Nurminen}}, \bibinfo {author} {\bibfnamefont {A.}~\bibnamefont {Kuronen}}, \
  and\ \bibinfo {author} {\bibfnamefont {K.}~\bibnamefont {Kaski}},\ }\href
  {\doibase 10.1103/PhysRevB.63.035407} {\bibfield  {journal} {\bibinfo
  {journal} {Phys. Rev. B}\ }\textbf {\bibinfo {volume} {63}},\ \bibinfo
  {pages} {035407} (\bibinfo {year} {2000})}\BibitemShut {NoStop}%
\bibitem [{\citenamefont {Lee}\ and\ \citenamefont
  {Barab\'asi}(1998)}]{Choongseop199873}%
  \BibitemOpen
  \bibfield  {author} {\bibinfo {author} {\bibfnamefont {C.}~\bibnamefont
  {Lee}}\ and\ \bibinfo {author} {\bibfnamefont {A.-L.}\ \bibnamefont
  {Barab\'asi}},\ }\href {\doibase http://dx.doi.org/10.1063/1.122542}
  {\bibfield  {journal} {\bibinfo  {journal} {Appl. Phys. Lett.}\ }\textbf
  {\bibinfo {volume} {73}},\ \bibinfo {pages} {2651} (\bibinfo {year}
  {1998})}\BibitemShut {NoStop}%
\bibitem [{\citenamefont {Wang}\ \emph {et~al.}(2007)\citenamefont {Wang},
  \citenamefont {Zhong}, \citenamefont {Zhu}, \citenamefont {Kalischewski},
  \citenamefont {Dou}, \citenamefont {Wedeking}, \citenamefont {Wang},
  \citenamefont {Heuer}, \citenamefont {Fuchs}, \citenamefont {Erker},\ and\
  \citenamefont {Chi}}]{Wang200798}%
  \BibitemOpen
  \bibfield  {author} {\bibinfo {author} {\bibfnamefont {W.~C.}\ \bibnamefont
  {Wang}}, \bibinfo {author} {\bibfnamefont {D.~Y.}\ \bibnamefont {Zhong}},
  \bibinfo {author} {\bibfnamefont {J.}~\bibnamefont {Zhu}}, \bibinfo {author}
  {\bibfnamefont {F.}~\bibnamefont {Kalischewski}}, \bibinfo {author}
  {\bibfnamefont {R.~F.}\ \bibnamefont {Dou}}, \bibinfo {author} {\bibfnamefont
  {K.}~\bibnamefont {Wedeking}}, \bibinfo {author} {\bibfnamefont
  {Y.}~\bibnamefont {Wang}}, \bibinfo {author} {\bibfnamefont {A.}~\bibnamefont
  {Heuer}}, \bibinfo {author} {\bibfnamefont {H.}~\bibnamefont {Fuchs}},
  \bibinfo {author} {\bibfnamefont {G.}~\bibnamefont {Erker}}, \ and\ \bibinfo
  {author} {\bibfnamefont {L.~F.}\ \bibnamefont {Chi}},\ }\href {\doibase
  10.1103/PhysRevLett.98.225504} {\bibfield  {journal} {\bibinfo  {journal}
  {Phys. Rev. Lett.}\ }\textbf {\bibinfo {volume} {98}},\ \bibinfo {pages}
  {225504} (\bibinfo {year} {2007})}\BibitemShut {NoStop}%
\bibitem [{\citenamefont {Kalischewski}\ \emph {et~al.}(2008)\citenamefont
  {Kalischewski}, \citenamefont {Zhu},\ and\ \citenamefont
  {Heuer}}]{Kalischewski200878}%
  \BibitemOpen
  \bibfield  {author} {\bibinfo {author} {\bibfnamefont {F.}~\bibnamefont
  {Kalischewski}}, \bibinfo {author} {\bibfnamefont {J.}~\bibnamefont {Zhu}}, \
  and\ \bibinfo {author} {\bibfnamefont {A.}~\bibnamefont {Heuer}},\ }\href
  {\doibase 10.1103/PhysRevB.78.155401} {\bibfield  {journal} {\bibinfo
  {journal} {Phys. Rev. B}\ }\textbf {\bibinfo {volume} {78}},\ \bibinfo
  {pages} {155401} (\bibinfo {year} {2008})}\BibitemShut {NoStop}%
\bibitem [{\citenamefont {Hu}\ \emph {et~al.}(2008)\citenamefont {Hu},
  \citenamefont {Gao},\ and\ \citenamefont {Liu}}]{Hu2008101}%
  \BibitemOpen
  \bibfield  {author} {\bibinfo {author} {\bibfnamefont {H.}~\bibnamefont
  {Hu}}, \bibinfo {author} {\bibfnamefont {H.~J.}\ \bibnamefont {Gao}}, \ and\
  \bibinfo {author} {\bibfnamefont {F.}~\bibnamefont {Liu}},\ }\href {\doibase
  10.1103/PhysRevLett.101.216102} {\bibfield  {journal} {\bibinfo  {journal}
  {Phys. Rev. Lett.}\ }\textbf {\bibinfo {volume} {101}},\ \bibinfo {pages}
  {216102} (\bibinfo {year} {2008})}\BibitemShut {NoStop}%
\bibitem [{\citenamefont {Engstfeld}\ \emph {et~al.}(2012)\citenamefont
  {Engstfeld}, \citenamefont {Hoster}, \citenamefont {Behm}, \citenamefont
  {Roelofs}, \citenamefont {Liu}, \citenamefont {Wang}, \citenamefont {Han},\
  and\ \citenamefont {Evans}}]{Engstfeld2012directed}%
  \BibitemOpen
  \bibfield  {author} {\bibinfo {author} {\bibfnamefont {A.~K.}\ \bibnamefont
  {Engstfeld}}, \bibinfo {author} {\bibfnamefont {H.~E.}\ \bibnamefont
  {Hoster}}, \bibinfo {author} {\bibfnamefont {R.~J.}\ \bibnamefont {Behm}},
  \bibinfo {author} {\bibfnamefont {L.~D.}\ \bibnamefont {Roelofs}}, \bibinfo
  {author} {\bibfnamefont {X.}~\bibnamefont {Liu}}, \bibinfo {author}
  {\bibfnamefont {C.-Z.}\ \bibnamefont {Wang}}, \bibinfo {author}
  {\bibfnamefont {Y.}~\bibnamefont {Han}}, \ and\ \bibinfo {author}
  {\bibfnamefont {J.~W.}\ \bibnamefont {Evans}},\ }\href {\doibase
  10.1103/PhysRevB.86.085442} {\bibfield  {journal} {\bibinfo  {journal} {Phys.
  Rev. B}\ }\textbf {\bibinfo {volume} {86}},\ \bibinfo {pages} {085442}
  (\bibinfo {year} {2012})}\BibitemShut {NoStop}%
\bibitem [{\citenamefont {N'Diaye}\ \emph {et~al.}(2006)\citenamefont
  {N'Diaye}, \citenamefont {Bleikamp}, \citenamefont {Feibelman},\ and\
  \citenamefont {Michely}}]{Diaye2006twidimensional}%
  \BibitemOpen
  \bibfield  {author} {\bibinfo {author} {\bibfnamefont {A.~T.}\ \bibnamefont
  {N'Diaye}}, \bibinfo {author} {\bibfnamefont {S.}~\bibnamefont {Bleikamp}},
  \bibinfo {author} {\bibfnamefont {P.~J.}\ \bibnamefont {Feibelman}}, \ and\
  \bibinfo {author} {\bibfnamefont {T.}~\bibnamefont {Michely}},\ }\href
  {\doibase 10.1103/PhysRevLett.97.215501} {\bibfield  {journal} {\bibinfo
  {journal} {Phys. Rev. Lett.}\ }\textbf {\bibinfo {volume} {97}},\ \bibinfo
  {pages} {215501} (\bibinfo {year} {2006})}\BibitemShut {NoStop}%
\bibitem [{\citenamefont {Gadzuk}(2008)}]{Gadzuk2008fundamental}%
  \BibitemOpen
  \bibfield  {author} {\bibinfo {author} {\bibfnamefont {J.}~\bibnamefont
  {Gadzuk}},\ }in\ \href@noop {} {\emph {\bibinfo {booktitle} {Dynamics}}},\
  \bibinfo {series and number} {Handbook of Surface Science},\ \bibinfo
  {editor} {edited by\ \bibinfo {editor} {\bibfnamefont {E.}~\bibnamefont
  {Hasselbrink}}\ and\ \bibinfo {editor} {\bibfnamefont {B.}~\bibnamefont
  {Lundqvist}}}\ (\bibinfo  {publisher} {Elsevier},\ \bibinfo {year}
  {2008})\BibitemShut {NoStop}%
\bibitem [{\citenamefont {Kleyn}(2008)}]{Kleyn2008basic}%
  \BibitemOpen
  \bibfield  {author} {\bibinfo {author} {\bibfnamefont {A.}~\bibnamefont
  {Kleyn}},\ }in\ \href@noop {} {\emph {\bibinfo {booktitle} {Dynamics}}},\
  \bibinfo {series and number} {Handbook of Surface Science},\ \bibinfo
  {editor} {edited by\ \bibinfo {editor} {\bibfnamefont {E.}~\bibnamefont
  {Hasselbrink}}\ and\ \bibinfo {editor} {\bibfnamefont {B.}~\bibnamefont
  {Lundqvist}}}\ (\bibinfo  {publisher} {Elsevier},\ \bibinfo {year}
  {2008})\BibitemShut {NoStop}%
\bibitem [{\citenamefont {Venables}(1973)}]{Venables1973rate}%
  \BibitemOpen
  \bibfield  {author} {\bibinfo {author} {\bibfnamefont {J.~A.}\ \bibnamefont
  {Venables}},\ }\href {\doibase 10.1080/14786437308219242} {\bibfield
  {journal} {\bibinfo  {journal} {Philos. Mag.}\ }\textbf {\bibinfo {volume}
  {27}},\ \bibinfo {pages} {697} (\bibinfo {year} {1973})}\BibitemShut
  {NoStop}%
\bibitem [{\citenamefont {Venables}\ \emph {et~al.}(1984)\citenamefont
  {Venables}, \citenamefont {Spiller},\ and\ \citenamefont
  {Hanbucken}}]{Hanbuecken1984399}%
  \BibitemOpen
  \bibfield  {author} {\bibinfo {author} {\bibfnamefont {J.~A.}\ \bibnamefont
  {Venables}}, \bibinfo {author} {\bibfnamefont {G.~D.~T.}\ \bibnamefont
  {Spiller}}, \ and\ \bibinfo {author} {\bibfnamefont {M.}~\bibnamefont
  {Hanbucken}},\ }\href@noop {} {\bibfield  {journal} {\bibinfo  {journal}
  {Rep. Prog. Phys.}\ }\textbf {\bibinfo {volume} {47}},\ \bibinfo {pages}
  {399} (\bibinfo {year} {1984})}\BibitemShut {NoStop}%
\bibitem [{\citenamefont {Kotrla}(1996)}]{Kotrla1996numerical}%
  \BibitemOpen
  \bibfield  {author} {\bibinfo {author} {\bibfnamefont {M.}~\bibnamefont
  {Kotrla}},\ }\href {\doibase http://dx.doi.org/10.1016/0010-4655(96)00023-9}
  {\bibfield  {journal} {\bibinfo  {journal} {Comput. Phys. Commun.}\ }\textbf
  {\bibinfo {volume} {97}},\ \bibinfo {pages} {82 } (\bibinfo {year}
  {1996})}\BibitemShut {NoStop}%
\bibitem [{\citenamefont {Levi}\ and\ \citenamefont
  {Kotrla}(1997)}]{levi97:_theor}%
  \BibitemOpen
  \bibfield  {author} {\bibinfo {author} {\bibfnamefont {A.~C.}\ \bibnamefont
  {Levi}}\ and\ \bibinfo {author} {\bibfnamefont {M.}~\bibnamefont {Kotrla}},\
  }\href {http://stacks.iop.org/0953-8984/9/i=2/a=001} {\bibfield  {journal}
  {\bibinfo  {journal} {J. Phys.- Condens. Mat.}\ }\textbf {\bibinfo {volume}
  {9}},\ \bibinfo {pages} {299} (\bibinfo {year} {1997})}\BibitemShut {NoStop}%
\bibitem [{\citenamefont {J\'onsson}(2000)}]{Jonsson2000theoretical}%
  \BibitemOpen
  \bibfield  {author} {\bibinfo {author} {\bibfnamefont {H.}~\bibnamefont
  {J\'onsson}},\ }\href {\doibase 10.1146/annurev.physchem.51.1.623} {\bibfield
   {journal} {\bibinfo  {journal} {Annu. Rev. Phys. Chem.}\ }\textbf {\bibinfo
  {volume} {51}},\ \bibinfo {pages} {623} (\bibinfo {year} {2000})},\ \bibinfo
  {note} {pMID: 11031295}\BibitemShut {NoStop}%
\bibitem [{\citenamefont {Voter}(2007)}]{Voter2007}%
  \BibitemOpen
  \bibfield  {author} {\bibinfo {author} {\bibfnamefont {A.}~\bibnamefont
  {Voter}},\ }in\ \href {\doibase 10.1007/978-1-4020-5295-8_1} {\emph {\bibinfo
  {booktitle} {Radiation Effects in Solids}}},\ \bibinfo {series} {NATO Science
  Series}, Vol.\ \bibinfo {volume} {235},\ \bibinfo {editor} {edited by\
  \bibinfo {editor} {\bibfnamefont {K.}~\bibnamefont {Sickafus}}, \bibinfo
  {editor} {\bibfnamefont {E.}~\bibnamefont {Kotomin}}, \ and\ \bibinfo
  {editor} {\bibfnamefont {B.}~\bibnamefont {Uberuaga}}}\ (\bibinfo
  {publisher} {Springer Netherlands},\ \bibinfo {year} {2007})\ pp.\ \bibinfo
  {pages} {1--23}\BibitemShut {NoStop}%
\bibitem [{\citenamefont {Petersen}\ \emph {et~al.}(2001)\citenamefont
  {Petersen}, \citenamefont {Ratsch}, \citenamefont {Caflisch},\ and\
  \citenamefont {Zangwill}}]{Petersen2001level}%
  \BibitemOpen
  \bibfield  {author} {\bibinfo {author} {\bibfnamefont {M.}~\bibnamefont
  {Petersen}}, \bibinfo {author} {\bibfnamefont {C.}~\bibnamefont {Ratsch}},
  \bibinfo {author} {\bibfnamefont {R.~E.}\ \bibnamefont {Caflisch}}, \ and\
  \bibinfo {author} {\bibfnamefont {A.}~\bibnamefont {Zangwill}},\ }\href
  {\doibase 10.1103/PhysRevE.64.061602} {\bibfield  {journal} {\bibinfo
  {journal} {Phys. Rev. E}\ }\textbf {\bibinfo {volume} {64}},\ \bibinfo
  {pages} {061602} (\bibinfo {year} {2001})}\BibitemShut {NoStop}%
\bibitem [{\citenamefont {Ratsch}\ \emph {et~al.}(2002)\citenamefont {Ratsch},
  \citenamefont {Gyure}, \citenamefont {Caflisch}, \citenamefont {Gibou},
  \citenamefont {Petersen}, \citenamefont {Kang}, \citenamefont {Garcia},\ and\
  \citenamefont {Vvedensky}}]{Ratsch2002level}%
  \BibitemOpen
  \bibfield  {author} {\bibinfo {author} {\bibfnamefont {C.}~\bibnamefont
  {Ratsch}}, \bibinfo {author} {\bibfnamefont {M.~F.}\ \bibnamefont {Gyure}},
  \bibinfo {author} {\bibfnamefont {R.~E.}\ \bibnamefont {Caflisch}}, \bibinfo
  {author} {\bibfnamefont {F.}~\bibnamefont {Gibou}}, \bibinfo {author}
  {\bibfnamefont {M.}~\bibnamefont {Petersen}}, \bibinfo {author}
  {\bibfnamefont {M.}~\bibnamefont {Kang}}, \bibinfo {author} {\bibfnamefont
  {J.}~\bibnamefont {Garcia}}, \ and\ \bibinfo {author} {\bibfnamefont {D.~D.}\
  \bibnamefont {Vvedensky}},\ }\href {\doibase 10.1103/PhysRevB.65.195403}
  {\bibfield  {journal} {\bibinfo  {journal} {Phys. Rev. B}\ }\textbf {\bibinfo
  {volume} {65}},\ \bibinfo {pages} {195403} (\bibinfo {year}
  {2002})}\BibitemShut {NoStop}%
\bibitem [{\citenamefont {Yu}\ and\ \citenamefont {Liu}(2004)}]{Yu2004phase}%
  \BibitemOpen
  \bibfield  {author} {\bibinfo {author} {\bibfnamefont {Y.-M.}\ \bibnamefont
  {Yu}}\ and\ \bibinfo {author} {\bibfnamefont {B.-G.}\ \bibnamefont {Liu}},\
  }\href {\doibase 10.1103/PhysRevE.69.021601} {\bibfield  {journal} {\bibinfo
  {journal} {Phys. Rev. E}\ }\textbf {\bibinfo {volume} {69}},\ \bibinfo
  {pages} {021601} (\bibinfo {year} {2004})}\BibitemShut {NoStop}%
\bibitem [{\citenamefont {Li}\ \emph {et~al.}(2003)\citenamefont {Li},
  \citenamefont {Bartelt},\ and\ \citenamefont {Evans}}]{Li2003geometry}%
  \BibitemOpen
  \bibfield  {author} {\bibinfo {author} {\bibfnamefont {M.}~\bibnamefont
  {Li}}, \bibinfo {author} {\bibfnamefont {M.~C.}\ \bibnamefont {Bartelt}}, \
  and\ \bibinfo {author} {\bibfnamefont {J.~W.}\ \bibnamefont {Evans}},\ }\href
  {\doibase 10.1103/PhysRevB.68.121401} {\bibfield  {journal} {\bibinfo
  {journal} {Phys. Rev. B}\ }\textbf {\bibinfo {volume} {68}},\ \bibinfo
  {pages} {121401} (\bibinfo {year} {2003})}\BibitemShut {NoStop}%
\bibitem [{\citenamefont {Russo}\ \emph {et~al.}(2004)\citenamefont {Russo},
  \citenamefont {Sander},\ and\ \citenamefont
  {Smereka}}]{Russo2004quasicontinuum}%
  \BibitemOpen
  \bibfield  {author} {\bibinfo {author} {\bibfnamefont {G.}~\bibnamefont
  {Russo}}, \bibinfo {author} {\bibfnamefont {L.~M.}\ \bibnamefont {Sander}}, \
  and\ \bibinfo {author} {\bibfnamefont {P.}~\bibnamefont {Smereka}},\ }\href
  {\doibase 10.1103/PhysRevB.69.121406} {\bibfield  {journal} {\bibinfo
  {journal} {Phys. Rev. B}\ }\textbf {\bibinfo {volume} {69}},\ \bibinfo
  {pages} {121406} (\bibinfo {year} {2004})}\BibitemShut {NoStop}%
\bibitem [{\citenamefont {Wang}\ \emph {et~al.}(2016)\citenamefont {Wang},
  \citenamefont {Buller}, \citenamefont {Wang}, \citenamefont {Heuer},
  \citenamefont {Zhang}, \citenamefont {Fuchs},\ and\ \citenamefont
  {Chi}}]{paperZero}%
  \BibitemOpen
  \bibfield  {author} {\bibinfo {author} {\bibfnamefont {H.}~\bibnamefont
  {Wang}}, \bibinfo {author} {\bibfnamefont {O.}~\bibnamefont {Buller}},
  \bibinfo {author} {\bibfnamefont {W.}~\bibnamefont {Wang}}, \bibinfo {author}
  {\bibfnamefont {A.}~\bibnamefont {Heuer}}, \bibinfo {author} {\bibfnamefont
  {D.}~\bibnamefont {Zhang}}, \bibinfo {author} {\bibfnamefont
  {H.}~\bibnamefont {Fuchs}}, \ and\ \bibinfo {author} {\bibfnamefont
  {L.}~\bibnamefont {Chi}},\ }\href {\doibase 10.1088/1367-2630/18/5/053006}
  {\bibfield  {journal} {\bibinfo  {journal} {New J. Phys.}\ }\textbf {\bibinfo
  {volume} {18}},\ \bibinfo {pages} {053006} (\bibinfo {year}
  {2016})}\BibitemShut {NoStop}%
\bibitem [{\citenamefont {Ranguelov}\ \emph {et~al.}(2007)\citenamefont
  {Ranguelov}, \citenamefont {Altman},\ and\ \citenamefont
  {Markov}}]{Ranguelov200775}%
  \BibitemOpen
  \bibfield  {author} {\bibinfo {author} {\bibfnamefont {B.}~\bibnamefont
  {Ranguelov}}, \bibinfo {author} {\bibfnamefont {M.~S.}\ \bibnamefont
  {Altman}}, \ and\ \bibinfo {author} {\bibfnamefont {I.}~\bibnamefont
  {Markov}},\ }\href {\doibase 10.1103/PhysRevB.75.245419} {\bibfield
  {journal} {\bibinfo  {journal} {Phys. Rev. B}\ }\textbf {\bibinfo {volume}
  {75}},\ \bibinfo {pages} {245419} (\bibinfo {year} {2007})}\BibitemShut
  {NoStop}%
\bibitem [{\citenamefont {Forrest}(1997)}]{Forrest199797}%
  \BibitemOpen
  \bibfield  {author} {\bibinfo {author} {\bibfnamefont {S.~R.}\ \bibnamefont
  {Forrest}},\ }\href {\doibase 10.1021/cr941014o} {\bibfield  {journal}
  {\bibinfo  {journal} {Chem. Rev.}\ }\textbf {\bibinfo {volume} {97}},\
  \bibinfo {pages} {1793} (\bibinfo {year} {1997})}\BibitemShut {NoStop}%
\bibitem [{\citenamefont {Briseno}\ \emph {et~al.}(2005)\citenamefont
  {Briseno}, \citenamefont {Aizenberg}, \citenamefont {Han}, \citenamefont
  {Penkala}, \citenamefont {Moon}, \citenamefont {Lovinger}, \citenamefont
  {Kloc},\ and\ \citenamefont {Bao}}]{Briseno2005patterned}%
  \BibitemOpen
  \bibfield  {author} {\bibinfo {author} {\bibfnamefont {A.~L.}\ \bibnamefont
  {Briseno}}, \bibinfo {author} {\bibfnamefont {J.}~\bibnamefont {Aizenberg}},
  \bibinfo {author} {\bibfnamefont {Y.-J.}\ \bibnamefont {Han}}, \bibinfo
  {author} {\bibfnamefont {R.~A.}\ \bibnamefont {Penkala}}, \bibinfo {author}
  {\bibfnamefont {H.}~\bibnamefont {Moon}}, \bibinfo {author} {\bibfnamefont
  {A.~J.}\ \bibnamefont {Lovinger}}, \bibinfo {author} {\bibfnamefont
  {C.}~\bibnamefont {Kloc}}, \ and\ \bibinfo {author} {\bibfnamefont
  {Z.}~\bibnamefont {Bao}},\ }\href {\doibase 10.1021/ja052919u} {\bibfield
  {journal} {\bibinfo  {journal} {J. Am. Chem. Soc.}\ }\textbf {\bibinfo
  {volume} {127}},\ \bibinfo {pages} {12164} (\bibinfo {year}
  {2005})}\BibitemShut {NoStop}%
\bibitem [{\citenamefont {Sundar}\ \emph {et~al.}(2004)\citenamefont {Sundar},
  \citenamefont {Zaumseil}, \citenamefont {Podzorov}, \citenamefont {Menard},
  \citenamefont {Willett}, \citenamefont {Someya}, \citenamefont {Gershenson},\
  and\ \citenamefont {Rogers}}]{Sundar2004elastomeric}%
  \BibitemOpen
  \bibfield  {author} {\bibinfo {author} {\bibfnamefont {V.~C.}\ \bibnamefont
  {Sundar}}, \bibinfo {author} {\bibfnamefont {J.}~\bibnamefont {Zaumseil}},
  \bibinfo {author} {\bibfnamefont {V.}~\bibnamefont {Podzorov}}, \bibinfo
  {author} {\bibfnamefont {E.}~\bibnamefont {Menard}}, \bibinfo {author}
  {\bibfnamefont {R.~L.}\ \bibnamefont {Willett}}, \bibinfo {author}
  {\bibfnamefont {T.}~\bibnamefont {Someya}}, \bibinfo {author} {\bibfnamefont
  {M.~E.}\ \bibnamefont {Gershenson}}, \ and\ \bibinfo {author} {\bibfnamefont
  {J.~A.}\ \bibnamefont {Rogers}},\ }\href {\doibase 10.1126/science.1094196}
  {\bibfield  {journal} {\bibinfo  {journal} {Science}\ }\textbf {\bibinfo
  {volume} {303}},\ \bibinfo {pages} {1644} (\bibinfo {year}
  {2004})}\BibitemShut {NoStop}%
\bibitem [{\citenamefont {Lucas}\ \emph {et~al.}(2012)\citenamefont {Lucas},
  \citenamefont {Trigaud},\ and\ \citenamefont
  {Videlot-Ackermann}}]{Lucas201261}%
  \BibitemOpen
  \bibfield  {author} {\bibinfo {author} {\bibfnamefont {B.}~\bibnamefont
  {Lucas}}, \bibinfo {author} {\bibfnamefont {T.}~\bibnamefont {Trigaud}}, \
  and\ \bibinfo {author} {\bibfnamefont {C.}~\bibnamefont
  {Videlot-Ackermann}},\ }\href {\doibase 10.1002/pi.3213} {\bibfield
  {journal} {\bibinfo  {journal} {Polym. Int.}\ }\textbf {\bibinfo {volume}
  {61}},\ \bibinfo {pages} {374} (\bibinfo {year} {2012})}\BibitemShut
  {NoStop}%
\bibitem [{\citenamefont {Forsythe}\ \emph {et~al.}(1998)\citenamefont
  {Forsythe}, \citenamefont {Morton}, \citenamefont {Tang},\ and\ \citenamefont
  {Gao}}]{Forsythe199873}%
  \BibitemOpen
  \bibfield  {author} {\bibinfo {author} {\bibfnamefont {E.~W.}\ \bibnamefont
  {Forsythe}}, \bibinfo {author} {\bibfnamefont {D.~C.}\ \bibnamefont
  {Morton}}, \bibinfo {author} {\bibfnamefont {C.~W.}\ \bibnamefont {Tang}}, \
  and\ \bibinfo {author} {\bibfnamefont {Y.}~\bibnamefont {Gao}},\ }\href
  {\doibase http://dx.doi.org/10.1063/1.122173"} {\bibfield  {journal}
  {\bibinfo  {journal} {App. Phys. Lett.}\ }\textbf {\bibinfo {volume} {73}},\
  \bibinfo {pages} {1457} (\bibinfo {year} {1998})}\BibitemShut {NoStop}%
\bibitem [{\citenamefont {Biehl}(2005)}]{Biehl2005lattice}%
  \BibitemOpen
  \bibfield  {author} {\bibinfo {author} {\bibfnamefont {M.}~\bibnamefont
  {Biehl}},\ }\enquote {\bibinfo {title} {Lattice gas models and kinetic monte
  carlo simulations of epitaxial growth},}\ in\ \href {\doibase
  10.1007/3-7643-7343-1_1} {\emph {\bibinfo {booktitle} {Multiscale Modeling in
  Epitaxial Growth}}},\ \bibinfo {editor} {edited by\ \bibinfo
  {editor} {\bibfnamefont {A.}~\bibnamefont {Voigt}}}\ (\bibinfo  {publisher}
  {Birkh{\"a}user Basel},\ \bibinfo {address} {Basel},\ \bibinfo {year}
  {2005})\ pp.\ \bibinfo {pages} {3--18}\BibitemShut {NoStop}%
\bibitem [{\citenamefont {Strobel}\ \emph {et~al.}(2001)\citenamefont
  {Strobel}, \citenamefont {Heinig},\ and\ \citenamefont
  {M\"oller}}]{Strobel2001domain}%
  \BibitemOpen
  \bibfield  {author} {\bibinfo {author} {\bibfnamefont {M.}~\bibnamefont
  {Strobel}}, \bibinfo {author} {\bibfnamefont {K.-H.}\ \bibnamefont {Heinig}},
  \ and\ \bibinfo {author} {\bibfnamefont {W.}~\bibnamefont {M\"oller}},\
  }\href {\doibase 10.1103/PhysRevB.64.245422} {\bibfield  {journal} {\bibinfo
  {journal} {Phys. Rev. B}\ }\textbf {\bibinfo {volume} {64}},\ \bibinfo
  {pages} {245422} (\bibinfo {year} {2001})}\BibitemShut {NoStop}%
\bibitem [{\citenamefont {M{\"u}ller}\ \emph {et~al.}(2002)\citenamefont
  {M{\"u}ller}, \citenamefont {Heinig},\ and\ \citenamefont
  {Schmidt}}]{Mueller2002template}%
  \BibitemOpen
  \bibfield  {author} {\bibinfo {author} {\bibfnamefont {T.}~\bibnamefont
  {M{\"u}ller}}, \bibinfo {author} {\bibfnamefont {K.-H.}\ \bibnamefont
  {Heinig}}, \ and\ \bibinfo {author} {\bibfnamefont {B.}~\bibnamefont
  {Schmidt}},\ }\href {\doibase
  http://dx.doi.org/10.1016/S0928-4931(01)00465-9} {\bibfield  {journal}
  {\bibinfo  {journal} {Mater. Sci. Eng. C}\ }\textbf {\bibinfo {volume}
  {19}},\ \bibinfo {pages} {209 } (\bibinfo {year} {2002})}\BibitemShut
  {NoStop}%
\bibitem [{\citenamefont {Metropolis}\ \emph {et~al.}(1953)\citenamefont
  {Metropolis}, \citenamefont {Rosenbluth}, \citenamefont {Rosenbluth},
  \citenamefont {Teller},\ and\ \citenamefont {Teller}}]{Metropolis53}%
  \BibitemOpen
  \bibfield  {author} {\bibinfo {author} {\bibfnamefont {N.}~\bibnamefont
  {Metropolis}}, \bibinfo {author} {\bibfnamefont {A.~W.}\ \bibnamefont
  {Rosenbluth}}, \bibinfo {author} {\bibfnamefont {M.~N.}\ \bibnamefont
  {Rosenbluth}}, \bibinfo {author} {\bibfnamefont {A.~H.}\ \bibnamefont
  {Teller}}, \ and\ \bibinfo {author} {\bibfnamefont {E.}~\bibnamefont
  {Teller}},\ }\href {\doibase 10.1063/1.1699114} {\bibfield  {journal}
  {\bibinfo  {journal} {J. Chem. Phys.}\ }\textbf {\bibinfo {volume} {21}},\
  \bibinfo {pages} {1087} (\bibinfo {year} {1953})}\BibitemShut {NoStop}%
\bibitem [{\citenamefont {Tewes}\ \emph {et~al.}(2017)\citenamefont {Tewes},
  \citenamefont {Buller}, \citenamefont {Heuer}, \citenamefont {Thiele},\ and\
  \citenamefont {Gurevich}}]{Tewes2017comparing}%
  \BibitemOpen
  \bibfield  {author} {\bibinfo {author} {\bibfnamefont {W.}~\bibnamefont
  {Tewes}}, \bibinfo {author} {\bibfnamefont {O.}~\bibnamefont {Buller}},
  \bibinfo {author} {\bibfnamefont {A.}~\bibnamefont {Heuer}}, \bibinfo
  {author} {\bibfnamefont {U.}~\bibnamefont {Thiele}}, \ and\ \bibinfo {author}
  {\bibfnamefont {S.~V.}\ \bibnamefont {Gurevich}},\ }\href {\doibase
  10.1063/1.4977739} {\bibfield  {journal} {\bibinfo  {journal} {J. Chem.
  Phys.}\ }\textbf {\bibinfo {volume} {146}},\ \bibinfo {pages} {094704}
  (\bibinfo {year} {2017})}\BibitemShut {NoStop}%
\bibitem [{\citenamefont {Burton}\ \emph {et~al.}(1951)\citenamefont {Burton},
  \citenamefont {Cabrera},\ and\ \citenamefont {Frank}}]{Burton1951299}%
  \BibitemOpen
  \bibfield  {author} {\bibinfo {author} {\bibfnamefont {W.~K.}\ \bibnamefont
  {Burton}}, \bibinfo {author} {\bibfnamefont {N.}~\bibnamefont {Cabrera}}, \
  and\ \bibinfo {author} {\bibfnamefont {F.~C.}\ \bibnamefont {Frank}},\ }\href
  {\doibase 10.1098/rsta.1951.0006} {\bibfield  {journal} {\bibinfo  {journal}
  {Philos. T. Roy. Soc. A}\ }\textbf {\bibinfo {volume} {243}},\ \bibinfo
  {pages} {299} (\bibinfo {year} {1951})}\BibitemShut {NoStop}%
\bibitem [{\citenamefont {Myers-Beaghton}\ and\ \citenamefont
  {Vvedensky}(1991)}]{Vvedensky199144}%
  \BibitemOpen
  \bibfield  {author} {\bibinfo {author} {\bibfnamefont {A.~K.}\ \bibnamefont
  {Myers-Beaghton}}\ and\ \bibinfo {author} {\bibfnamefont {D.~D.}\
  \bibnamefont {Vvedensky}},\ }\href {\doibase 10.1103/PhysRevA.44.2457}
  {\bibfield  {journal} {\bibinfo  {journal} {Phys. Rev. A}\ }\textbf {\bibinfo
  {volume} {44}},\ \bibinfo {pages} {2457} (\bibinfo {year}
  {1991})}\BibitemShut {NoStop}%
\bibitem [{\citenamefont {Amar}\ and\ \citenamefont
  {Family}(1995)}]{Amar199574}%
  \BibitemOpen
  \bibfield  {author} {\bibinfo {author} {\bibfnamefont {J.~G.}\ \bibnamefont
  {Amar}}\ and\ \bibinfo {author} {\bibfnamefont {F.}~\bibnamefont {Family}},\
  }\href {\doibase 10.1103/PhysRevLett.74.2066} {\bibfield  {journal} {\bibinfo
   {journal} {Phys. Rev. Lett.}\ }\textbf {\bibinfo {volume} {74}},\ \bibinfo
  {pages} {2066} (\bibinfo {year} {1995})}\BibitemShut {NoStop}%
\end{thebibliography}
\end{document}